\documentclass[review]{elsarticle}

\usepackage{lineno,hyperref}
\modulolinenumbers[5]

\journal{Comp Math Meth Med}

\usepackage{verbatim}
\usepackage{amsmath} %for splits in equations
\usepackage{graphicx}
\usepackage{grffile}
\usepackage{tikz}
\usepackage{tikz-cd}
\usepackage{tkz-graph}
\usepackage{tkz-berge}
\usepackage{amsfonts}
\usepackage{listings}
\usepackage{fancyvrb}
\usepackage{latexml}
\usepackage{amssymb}

\usepackage{amsthm}
\newtheorem*{definition}{Definition}

\usepackage{url}
\usepackage{etoolbox} %http://tex.stackexchange.com/questions/86529/breqn-and-lineno-incompatibility
\usepackage{breqn}   %for better splits in equations

\BeforeBeginEnvironment{dmath}{\par\begin{nolinenumbers}}%
\AfterEndEnvironment{dmath}{\end{nolinenumbers}}%
\BeforeBeginEnvironment{dmath*}{\par\begin{nolinenumbers}}%
\AfterEndEnvironment{dmath*}{\end{nolinenumbers}}
\BeforeBeginEnvironment{dgroup*}{\par\begin{nolinenumbers}}%
\AfterEndEnvironment{dgroup*}{\end{nolinenumbers}}

\newcommand\commentout[1]{}

\newcommand{\hl}[1]{#1}

\newcommand{\cartprod}{\;\Box\;}

\begin{document}
\begin{frontmatter}

\title{Products of Compartmental Models in Epidemiology}

\author[proctor]{Lee Worden}

\author[proctor,ophth,biostat]{Travis~C.~Porco\corref{corrauth}}
\cortext[corrauth]{Corresponding author}
\ead{Travis.Porco@UCSF.edu}

\address[proctor]{Francis I. Proctor Foundation, University of California San Francisco, San Francisco, California, USA}
\address[ophth]{Department of Ophthalmology, University of California, San Francisco, California, USA}
\address[biostat]{Department of Epidemiology and Biostatistics, University of California San Francisco, San Francisco, California, USA}

\begin{abstract}
In this paper, we show that many structured epidemic models
may be described using a straightforward product structure.
Such products, derived from products of directed graphs,
may represent useful refinements including geographic and
demographic structure, age structure,
gender, risk groups, or immunity status.
Extension to multi-strain dynamics, i.e.\ pathogen heterogeneity,
is also shown to be feasible in this framework.
Systematic use of such products may aid in model development
and exploration, can yield insight, and
could form the basis of a systematic approach to
numerical structural sensitivity analysis.
\end{abstract}


\end{frontmatter}

%\linenumbers

\section{Introduction}

Simple epidemic models aim at insight through simplicity;
complex models aim at realism through detail \cite{aron1989}.
Both simple
and complex models are still being developed
(e.g., \cite{saadroy-shuai-vandendriessche2016,althaus2015,blumberg-worden-enanoria2015,liu-enanoria-zipprich2015,macal-north-collier2014,merler-ajelli-fumanelli2015}).
Addition of epidemiological refinements, such as 
age structure, gender, geographic separation, or pathogen strains
in general changes the behavior of simple models, and thus we must 
systematically compare models with different features.

In this paper, we show that many structured epidemic models
may be described using a straightforward
product structure.  Such products therefore provide a compact representation for
a family of related models and could facilitate model comparison
and structural sensitivity analysis.
Examples include modeling
host susceptibility groups, gender, age structure, multiple subtypes, and geographic separation.
Our attention will be restricted to compartmental models
\cite{jacquez1972compartmental,matis2012stochastic,anderson1991infectious},
focusing on mathematical epidemiology \cite{kermack1927contribution,bailey1975,hethcote1976,anderson-may1991,hethcote1994,hethcote2000,vandendriessche-watmough2002,martcheva2015}.

The product we describe is related to standard graph products.
The relation between compartmental models and graph theoretic or
network concepts has been long appreciated \cite{mason56,lewis77}, and moreover,
Markov processes arising on product spaces have been analyzed by
probabilists \cite{vasershtein69}.  The graph
structure arises when
dynamical variables will be represented as vertices of a graph, 
representing the number of individuals in a given compartment.
Individuals
may change state, such changes being represented by an arc from one
vertex to another, labeled with the instantaneous rate at which such a transition
would occur.  

\section{Motivating Example: Community-Structured Epidemic Model}

Consider a simple SI (susceptible to infective) model describing an epidemic with no
recovery.  Individuals transition from
susceptible to infective, and never return to the uninfected state.  The number
of infected individuals is denoted $I$; of susceptible individuals $S$.

\begin{figure}
\centering
% WMD file Watson-SI.crop.pdf
\includegraphics{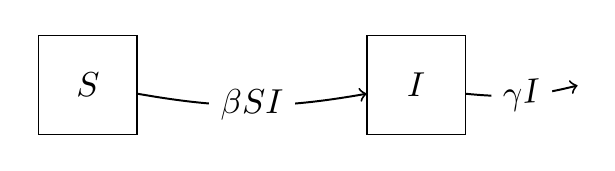}
\caption{ \label{fig:SI}
Directed graph diagram of simple SI model.
}
\end{figure}

This compartmental model is diagrammed in \hl{Figure~\mbox{\ref{fig:SI}}}.
The corresponding ODE system may be written
\begin{dgroup*}
\begin{dmath*}
 \frac{dS}{dt} = - \beta S I,
\end{dmath*}\begin{dmath*}
 \frac{dI}{dt} = \beta S I - \gamma I.
\end{dmath*}
\end{dgroup*}
%When $\gamma=0$,
%the SI system itself is the one dimensional
%${dI}/{dt} = \beta (N-I)I$, with
%analytic solution $I(t)={I_0}/\left({(N-I_0)e^{\beta N t} + I_0}\right)$.
Here,
$\beta$ is a transmission coefficient, and
$\gamma$ is the per capita mortality or removal rate due to disease.  In this model,
we ignore population birth and death due to other causes.

\begin{figure}
\centering
% WMD file Watson-combined.crop.pdf
\includegraphics{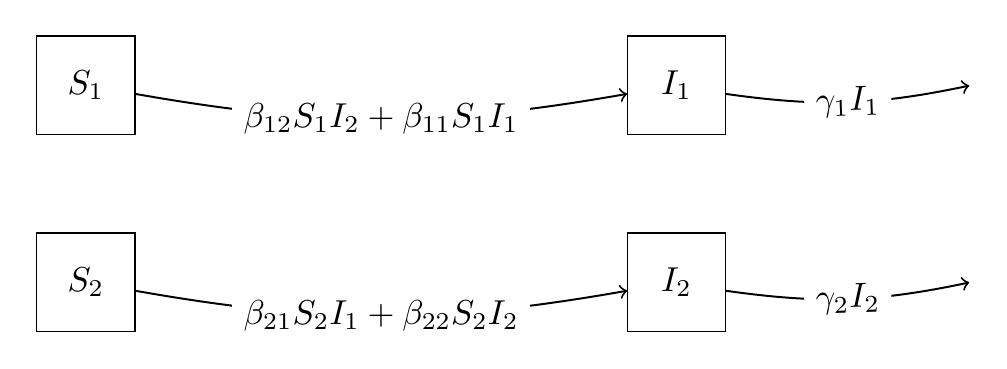}
\caption{ \label{fig:Watson-combined}
Directed graph diagram of
Watson model \cite{watson72} defined by adding
community structure to the SI model.
}
\end{figure}

A simple extension to include heterogeneous epidemic dynamics
in multiple communities
was introduced by Watson~\cite{rushton-mautner55,watson72}.
In this model,
no migration between communities is assumed.
However, individuals in one
community cause infection in other communities,
with the structure seen in \hl{Figure~\mbox{\ref{fig:Watson-combined}}}.
The equations are
\begin{dgroup*}
\begin{dmath*}
 \frac{dS_i}{dt} = - \sum_j \beta_{ij} S_i I_j,\quad{i=1,\ldots,n}
\end{dmath*}\begin{dmath*}
 \frac{dI_i}{dt} = \sum_j \beta_{ij} S_i I_j - \gamma_i I_i,\quad{i=1,\ldots,n}
\end{dmath*}
\end{dgroup*}
where $n$ is the number of communities modeled.
In the Watson model, in general the transmission coefficients may
differ when considering transmission to susceptibles in one community
from infectives in any community (whether the same or not).
Each community is additionally
assumed to have a different rate $\gamma_i$ of removal of infectives
due to mortality (or other causes), though
these can be assumed to be identical if desired.

This model extends the one-community
SI model, by structuring the population into
multiple communities.  In the following section, we will 
show that the structured model developed by Watson can
be straightforwardly defined as the product of
the single-community SI model and a model describing
community structure.  We will then illustrate other uses
of this product, including age-structure, gender, heterogeneity of risk, and
co-transmission of multiple diseases.

\section{ Graph products}

A directed graph is defined as a set of vertices,
each identified by a unique label,
together with a set of arrows, or arcs,
each connecting a source vertex to a target vertex.
In this paper we are concerned only with directed graphs,
not undirected ones.
A number of different products of directed graphs are defined,
two of which are relevant.

\subsection{Cartesian product}
Consider finite directed graphs $A$ and $B$, with $n_A$ and
$n_B$ vertices, respectively.  
The Cartesian product of these graphs \cite{imrich-klavzar2000}
is a graph $A\cartprod B$ whose vertex set is the set of ordered pairs
$(v,w)$ for all vertices $v$ of $A$ and $w$ of $B$
(that is, the Cartesian product of the vertex sets of the
factor graphs $A$ and $B$).
The arcs of $A\cartprod B$ consist of an arc from
$(v,w_s)$ to $(v,w_t)$ for every $v$, wherever there is an
arc from $w_s$ to $w_t$ in the factor graph $B$,
and an arc from $(v_s,w)$ to $(v_t,w)$ for every $w$ 
wherever there is an arc from $v_s$ to $v_t$ in $A$.

We can speak of ``levels'' in the sense that each vertex of a
factor model corresponds to a subset, or level, of vertices of
the product model.
The product replicates all the arcs of $B$ at every level of $A$,
and all the arcs of $A$ at every level of $B$.
Suppose we have two vertices $(a_i, b_j)$ and $(a_k, b_j)$,
whose second coordinate is the same, i.e. which map
to the same level of $B$;
it will be helpful to call these ``siblings'',
and to say they ``descend'' from a common ``factor vertex''
$B$; similarly for vertices with the same first coordinate.

\hl{Figure \mbox{\ref{fig:graph-products}}(a),~(b),~and~(c)}
illustrates two directed graphs and their Cartesian product,
respectively.

More generally, graphs with multiple arcs joining a pair of vertices
can be defined,
and the Cartesian product definition
above can be applied in this case as well.

\subsection{Strong product}

The strong product of two directed graphs $A$ and $B$
includes more arcs than the Cartesian product\cite{imrich-klavzar2000}.
This product $A\boxtimes B$ has the same vertex set,
the Cartesian product of the factors' vertex sets,
but in addition to the arcs of the Cartesian product graphs,
it also includes all arcs from $(v_s,w_s)$ to $(v_t,w_t)$
where there is an arc from $v_s$ to $v_t$ \emph{and} an arc
from $w_s$ to $w_t$.

\begin{figure}
\centering
(a) % WMD file G_A.crop.pdf
\includegraphics{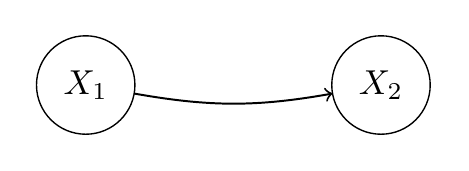}
\\
(b) % WMD file G_B.crop.pdf
\includegraphics{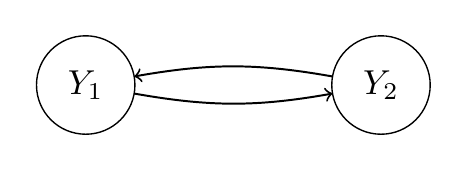}
\\
(c) % WMD file graph-Cartesian-product.crop.pdf
\includegraphics{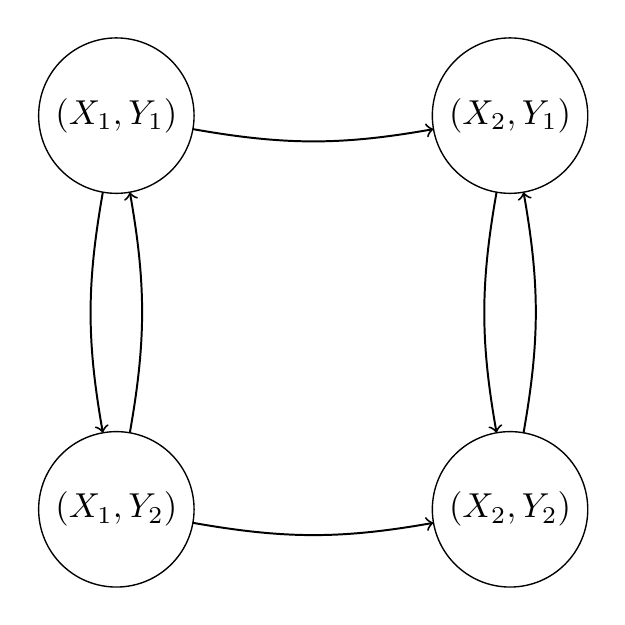}
\\
(d) % WMD file graph-strong-product.crop.pdf
\includegraphics{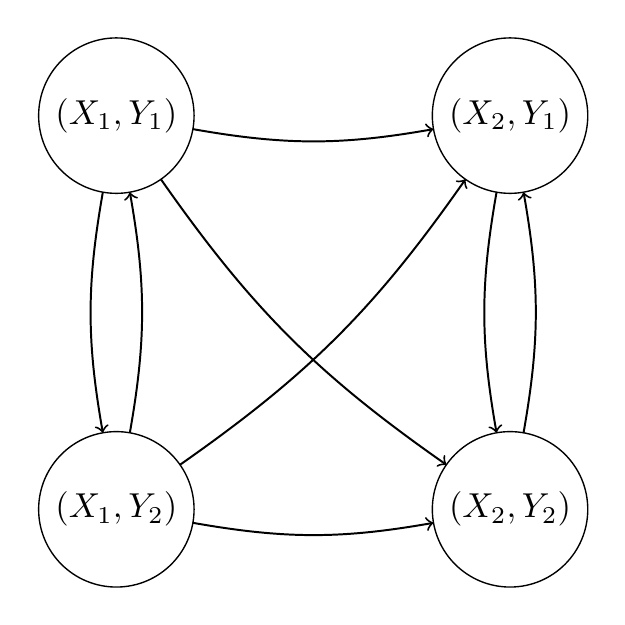}
\caption{ \label{fig:graph-products}
(a, b) Example directed graphs $A$ and $B$, respectively;
(c) Cartesian product $A\cartprod B$;
(d) strong product $A\boxtimes B$.
}
\end{figure}

\hl{Figure~\mbox{\ref{fig:graph-products}}(d)} illustrates
the strong product of the graphs of
\hl{Figure \mbox{\ref{fig:graph-products}}(a)~and~(b)}.

The symbols $\Box$ and $\boxtimes$ for these operations
are chosen to evoke the structure of the product graphs,
as illustrated in
\hl{Figure \mbox{\ref{fig:graph-products}}(c)~and~(d)}.
These graph products are discussed in more detail in the Appendix.

\section{Products of models}

\subsection{Linear compartmental models}

A compartmental model (whether in 
population biology, epidemiology, or pharmacology) is often represented by a
diagram such as in \hl{Figure~\mbox{\ref{fig:SI}}}, which has the form of 
a directed graph (formally, a directed multigraph) with labels on arcs.
Multiple arcs may connect
a single pair of compartments, representing multiple
processes influencing that transition with potentially
different rates.
As before, the vertices of the graph are compartments and its arcs are
transitions, with labels specifying the transition rates.
(We consider a compartmental model to be an abstract object
isomorphic to its directed-multigraph diagram.)
As is well known, a compartmental model diagram can
be represented by a system of ordinary differential equations or a 
continuous Markov jump process (among others).
(For instance, a compartmental
model with a single compartment $N$, with a single inflow with
rate $\Lambda$ and outflow with rate $\mu N$ can be represented by
the simple stochastic immigration-death process \cite{baileystoch}
or by
the elementary ordinary differential equation model $\frac{dN}{dt} = \Lambda - \mu N$.)

The class of linear compartmental models we consider in this section include the
ordinary differential equation models of the form
\[
\frac{dX}{dt} = a + M X,
\]
where $X$ is a vector of $n$ state variables, $a$ is a vector of constant inflows, and $M$ is an
$n \times n$ transition rate matrix.
The general compartmental model, with sources and sink terms, can be represented in the same graphical way by
considering special source and sink vertices in the graph.  

A Cartesian product of linear compartmental models will be defined in a way that is similar
to the Cartesian product of graphs.  
Suppose $A_1$, $A_2$, $\ldots$, $A_K$ be
the states in model $A$; let $B_1, \ldots, B_L$ be the states of model $B$.  The Cartesian
product of $A$ with $B$ will have states $(A_i, B_j)$ with $i = 1, \ldots, K$ and $j=1,\ldots, L$.
Two states $(A_i,B_j)$ and $(A_i,B_{j'})$ are {\it siblings} in the same {\it level} $A_i$ of the product.
If we began, for example, with an epidemic model with states susceptible, infective, and
removed (SIR), and wished to construct a product with a geographic model of multiple regions, we
would expect to have susceptibles, infectives, and removed individuals in each region.  

The Cartesian product of graphs, as we saw, replicates each arc of each factor graph for each
vertex of the other factor graph.  In a compartmental model of a population system, this would
correspond to the very common assumption of competing independent exponential risks.  For example,
consider once again a simple SIR epidemic model, with infection and recovery, and a model of
two communities with migration between them.  In a Cartesian product of the two, we may wish
to allow infection and recovery within each community as well as migration of susceptibles from one
community to another, migration of infectives, and migration of recovered individuals.  In the product
model, infectives in one community, for example, should be able to move to the other community or
recover within their own community---a feature exactly reflected in the structure of a Cartesian graph product.  

However, note that in general,
we may well wish to assume differences in these parameters.  We may wish to assume, for example, that
recovery rates are higher in one community, or that migration rates of infectives are lower than for
susceptibles.  Unlike a Cartesian graph product, a Cartesian product of compartmental models must
take into account the arc labels, which are the transition rates; in general, new parameters are
necessarily introduced.

We propose the following definition for a Cartesian product of linear compartmental models.
If a transition in model $B$ from $B_j$ to $B_{j'}$ occurs with rate $\gamma$, then for every
state $i$ in model $A$, a transition in the product model occurs from
$(A_i,B_j)$ to $(A_i,B_{j'})$ at rate $\gamma_i$.  Similarly,
if a transition in model $A$ from $A_i$ to $A_{i'}$ occurs with rate $\theta$, then for every
state $j$ in model $B$, a transition in the product model occurs from
$(A_i,B_j)$ to $(A_{i'},B_j)$ at rate $\theta_j$.  

The presence of sources and sinks does not add any fundamental complications.
If a transition in model $B$ from $B_j$ to a sink occurs with rate $\mu$, then for every state $i$ in model $A$,
a transition from $(A_i,B_j)$ to the sink occurs with rate $\mu_i$ (similarly, {\it mutatis mutandis}, for
transitions in $A$ to a sink).
Finally, if a transition from a source to state $B_j$ in model $B$ occurs at rate $\Lambda$, then in the product model,
for every state $i$ in model $A$, a transition from a source to $(A_i,B_j)$ occurs at rate $\Lambda_i$ (and
similarly for transitions from the source which appear in model $A$).  These, and only these, transitions constitute
the product model.  

See the Appendix for more detail on the
Cartesian product of linear models.

%Let $U_1=(A_i,B_j)$ and $U_2=(A_{i'},B_{j'})$ be two states in the product model.  Then
%a transition from $U_1$ to $U_2$ occurs if and only if one of the following is true:
%\begin{itemize}
%\item $i=i'$ and a transition from $B_j$ to $B_{j'}$ occurs in factor model $B$.
%\item $j=j'$ and a transition from $A_i$ to $A_{i'}$ occurs in factor model $A$.
%\end{itemize}
%A transition from $U_1$ to a sink occurs if and only if one of the following is true:
%\begin{itemize}
%\item A transition from $B_j$ to a sink occurs in factor model $B$.
%\item A transition from $A_i$ to a sink occurs in factor model $A$.
%\end{itemize}
%A transition from a source to $U_2$ occurs if and only if one of the following is true:
%\begin{itemize}
%\item A transition from a source to $B_{j'}$ occurs in factor model $B$.
%\item A transition from a source to $A_i$ occurs in factor model $A$.
%\end{itemize}

\section{Epidemic models}
How can products like the Cartesian and strong products of graphs
be used in formulating epidemic models?  As we shall see,
the product reviewed above can be extended to this case as well.  We must
extend the Cartesian product of compartmental models to allow
interaction between different populations.  (We note that similar
considerations apply in the more general ecological modeling
setting, including Lotka-Volterra predator-prey and competition
equations, but we will not pursue these applications.)

\subsection{Structured SI model}

In this section, we return to the classical Watson epidemic model,
representing the SI epidemic in multiple regions.  We will extend the Cartesian product of 
linear compartmental models, showing that the Watson model is a product of
the simple SI model and a geographic model.  In this special case, the geographic
model will have no transitions at all.

The SI model of the transmission process
is the one discussed above, with two states
$S$ and $I$, and transitions as pictured in \hl{Figure~\mbox{\ref{fig:SI}}}.

\begin{figure}
\centering
(a) % WMD file N12.crop.pdf
\includegraphics{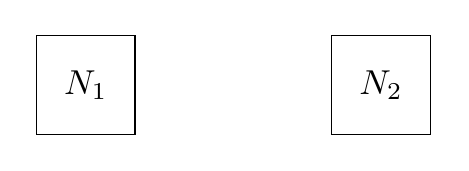}
\\
(b) % WMD file SI12lambda.crop.pdf
\includegraphics{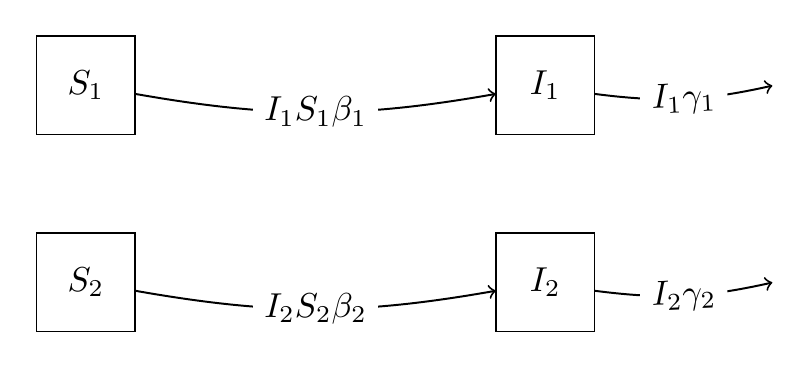}
\caption{ \label{fig:SI-simple}
(a) Factor model representing community structure with no migration;
(b) simple Cartesian product formula applied
to the SI and community structure models,
giving an incorrect result.
}
\end{figure}

We now define a factor model which distinguishes
individuals by community, to be combined with the SI model.
If there
are $n$ communities, let $N_1$, $N_2$, $\ldots$, $N_n$ be the number of
individuals in each community.  If no migration takes place, and we
ignore demographic turnover, this model corresponds to the differential
equation system ${dN_k}/{dt}=0$ for all $k$.
For simplicity, we will illustrate only the $n=2$ case
(\hl{Figure~\mbox{\ref{fig:SI-simple}(a)}}).

\begin{figure}
\centering
% WMD file Watson.crop.pdf
\includegraphics{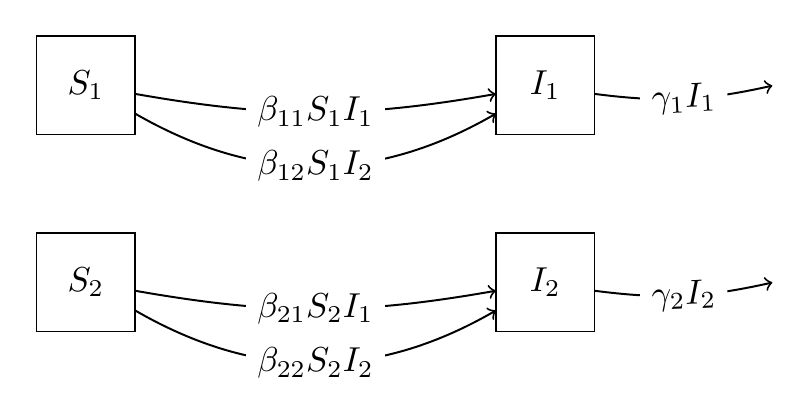}
\caption{ \label{fig:Watson}
Directed multigraph diagram of
Watson model \cite{watson72} defined by applying
the extended Cartesian product operation to the SI
and two-community models.
}
\end{figure}

The state space of the product model
will consist of the numbers of susceptibles and infectives
in community 1 and community 2, 
ordered pairs such as $(S,N_1)$, which can be given names
$S_1$, $I_1$, $S_2$, and $I_2$.
Naive application of the Cartesian product for compartmental
models would begin with the observation that
the SI factor model includes a transition from $S$ to
$I$ at rate $\beta I$.  We would then iterate over
the levels $j=1, 2$ of the community model.  We need a transition
from $(S,N_1)\equiv S_1$ to $(I,N_1)\equiv I_1$, but at
what rate?  
Generalizing the Cartesian product formula given above
in the most direct way produces a model with transition
rates $\beta_1 S_1 I_1$ and $\beta_2 S_2 I_2$,
corresponding to the graph seen in \hl{Figure~\mbox{\ref{fig:SI-simple}(b)}}.
This is a technically valid compartmental model, but
it does not account for potential transmission between
infectives in one community (e.g.\ $I_2$)
and susceptibles in the other (e.g.\ $S_1$).

We must therefore extend the Cartesian product of 
compartmental models.  
In this example, we must take into account that the rate of transmission
between a susceptible and an infective individual
depends on the community membership of the infective as well
as that of the susceptible.
The extended definition is as follows.

As above, let $A$ and $B$ be
models.  The state space for the extended Cartesian product $C$ is, as
before, the Cartesian product of the state spaces of $A$ and of $B$.  
For $A$, the transition rates include functional forms
$f(A_i,A_j)$, i.e. functional dependencies on one or more
states of $A$.

In the example of the Watson epidemic model, we have the following.
The transition rate denoting transmission events in the SI model
has rate $f(S,I)=\beta S I$.
We will construct the product model using the rule that
from every compartment $S_i\in\{S_1,S_2\}$, that is,
for every compartment descended from the $S$ compartment,
there is a transition to its corresponding sibling descended
from the $I$ compartment, at rate
$f_{ij}(S_i,I_j)=\beta_{ij}S_iI_j$,
for every $I_j\in\{I_1,I_2\}$.
Note that the infective compartment $I_j$ in this definition
is distinct from the target vertex of the transition ---
the transition arc points from $S_i$ to $I_i$,
on the level $i$ of the source vertex,
but the infective compartment $I_j$ ranges over levels $j$
of the product model independently of the source ---
and this distinction is crucial to defining the correct
set of transitions.

%Above, we have represented the transition rate matrix as a sum of arcs,
%e.g.\ $M_A=\sum_{\alpha}M_A^{\alpha}$,
%leading to a Cartesian product
%\[
%M_C = \sum_{\alpha} 
%\Lambda^{\alpha}_B \otimes M_A^{\alpha} + \sum_{\beta} M_B^{\beta} \otimes \Lambda^{\beta}_A,
%\]
%but now, each arc may involve a functional dependency on
%a set of state variables that must be handled explicitly.
%In this example the above expansion by arcs is
%\begin{dgroup*}
%\begin{dmath*}
%M_C = \Lambda^f_B \otimes M_A^f
%\end{dmath*}
%\begin{dmath*}
% = \left[\begin{array}{cccc}
%    - \lambda_1 f(I) & 0 & 0 & 0 \\
%      \lambda_1 f(I) & 0 & 0 & 0 \\
%    0 & 0 & - \lambda_2 f(I) & 0 \\
%    0 & 0 &   \lambda_2 f(I) & 0
%  \end{array}\right]
%\end{dmath*}
%\end{dgroup*}
%Instead, we need
%\begin{dgroup*}
%\begin{dmath*}
%M_C = \sum_j \Lambda^{f,j}_B \otimes M_A^f(I_j)
%\end{dmath*}
%\begin{dmath*}
%= \left[\begin{array}{cccc}
%    - \lambda_{11} f(I_1) - \lambda_{12} f(I_2) & 0 & 0 & 0 \\
%      \lambda_{11} f(I_1) + \lambda_{12} f(I_2) & 0 & 0 & 0 \\
%    0 & 0 & - \lambda_{21} f(I_1) - \lambda_{22} f(I_2) & 0 \\
%    0 & 0 &   \lambda_{21} f(I_1) + \lambda_{22} f(I_2) & 0
%  \end{array}\right].
%\end{dmath*}
%\end{dgroup*}

Where our earlier definition constructs one arc from each
$S$ compartment to its corresponding $I$ compartment,
this definition constructs one arc from each $S$ compartment
to its $I$ compartment \emph{for each}
infective compartment that can transmit to those susceptibles.
%That is, an arc appears from state $S_i$ to $I_i$
%in the product, with total transition value given by
%$f_{ij}(S_i,I_j)=\beta_{ij} S_i I_j$, for every $S_i \in {S_1,S_2}$, 
%and $I_j \in {I_1,I_2}$, as in this table:
%\[
%\begin{array}{cccc}
%S_i   & I_j   & \text{total rate} \\ \hline
%S_1   & I_1   & \beta_{11} S_1 I_1 \\
%S_1   & I_2   & \beta_{12} S_1 I_2 \\
%S_2   & I_1   & \beta_{21} S_2 I_1 \\
%S_2   & I_2   & \beta_{22} S_2 I_2 \\
%\end{array}
%\]
%In the table, we condense the notation for $\beta$ to reflect only the 
%relevant indices.
This yields the model shown in \hl{Figure~\mbox{\ref{fig:Watson}}}.
This extended Cartesian product yields two arcs for transitions from $S_1$ to $I_1$,
the first reflecting our intent that individuals in community 1 can cause infections in their
own community, and the second reflecting transmission to community 1 from community 2.  This
can be canonically represented as a single arc whose rate is the sum of the rates in the
individual arcs, which in this example, is $(\beta_{11} I_1 + \beta_{12} I_2) S_1$, as in
the original presentation of this model by Watson \cite{watson72}.
Similarly, two arcs appear for transitions from $S_2$ to $I_2$.  Thus, the extended
Cartesian product correctly represents the Watson model as a product of a within-community
epidemic process and a geographic model.

Here we provide a formal definition of this product:

\begin{definition}  A simple Cartesian product of two
compartmental models $A$ and $B$ is a compartmental
model $A\cartprod B$ whose set of compartments is
the set of ordered pairs $(X_i,Y_j)$ for every compartment $X_i$
of $A$ and $Y_j$ of $B$.

For every arc $\alpha$ of
model $A$, with per capita transition rate
$f^{\alpha}(X_s,Z_1,\ldots,Z_k)$,
source compartment $X_s$, and target compartment $X_t$,
the arcs of the product model include all arcs of the form
\[ x_s \xrightarrow{f^{\alpha}_{i,\ldots}(x_s,z_{1},\ldots,z_{k})} x_t \]
where $x_s,z_1,\ldots,$ and $z_k$ 
range over all compartments of the product model descending from
$X_s,Z_1,\ldots,$ and $Z_k$ respectively,
and $x_t$ is the compartment descending from $X_t$ that is
otherwise on the same level as $x_s$;
together with the corresponding arcs derived from the arcs of
factor model $B$.
The subscripts of $f^{\alpha}_{i,\ldots}$ distinguish the different arcs
by providing the names of the levels to which all of the product compartments
$x_s,z_1,\ldots,z_k$ belong.
The set of arcs of the product model consists of only the above arcs.
\end{definition}

The product transition rates
$f^{\alpha}_{i,\ldots}(x_s,z_1,\ldots,z_k)$
can be defined as needed, to generate an appropriately concise
form for the transition rate functions,
set unneeded transition rates to zero, or to do
other work of specifying the details of the combined epidemic
dynamics.
Examples below demonstrate several ways of using these functions
to construct specific models.

Before introducing a series of examples of product models with
epidemiological application,
we note that the product compartments,
defined as ordered $n$-tuples such as $(S,N_1)$,
can be assigned variable names such as $S_1$
in a number of ways. We will use several different
naming conventions in our examples.
Likewise the parameters such as $\beta$ and $\gamma$
need to be mapped in product transitions
to differentiated variables such as $\beta_{12}$,
$\gamma_{1}$, etc., as appropriate to the application.
We consider this to be part of the definition of the
function $f^{\alpha}_{ij}(S_i,I_j)$ and other
rate functions.

\subsection{Community model featuring demographics}

\begin{figure}
\centering
% WMD file demo-product.boxes.crop.pdf
\includegraphics{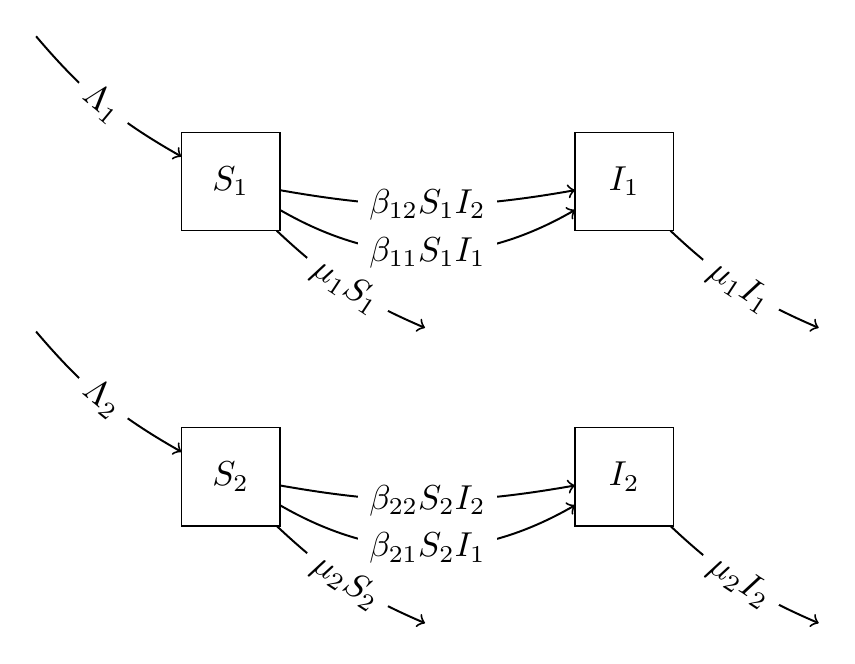}

\caption{ \label{fig:demo-product}
Product model of SI process in two communities.
}
\end{figure}

We note that the disease process factor model may be
generalized to include
demographic turnover (``vital dynamics'').  For example, this may feature
a constant inflow of new susceptibles and an exponential mortality or removal,
the SI model could be expressed in the form
\begin{dgroup*}
\begin{dmath*}
 \frac{dS}{dt} = \Lambda -\beta S I - \mu S
\end{dmath*}\begin{dmath*}
  \frac{dI}{dt} = \beta S I - \mu I
\end{dmath*}
\end{dgroup*}
Here, $\Lambda$ is a constant recruitment rate, and $\mu$ a per-capita
death rate (see, for example, \cite{anderson-may1991}).
If sources and sinks are considered to be special compartments,
the above definition encompasses such inflow and outflow transitions.
If we construct the
extended Cartesian product model of this SI process with the same community
model, we obtain the correct product model, with differential equations
\begin{dgroup*}
\begin{dmath*}
\frac{dS_i}{dt} = \Lambda_i - \sum_j \beta_{ij} S_i I_j - \mu_i S_i
\end{dmath*}
\begin{dmath*}
\frac{dI_i}{dt} =  \sum_j \beta_{ij} S_i I_j - \mu_i I_i.
\end{dmath*}
\end{dgroup*}
This model is illustrated in \hl{Figure~\mbox{\ref{fig:demo-product}}}.

%Note that instead, if we preserve the original form of the SI model, without vital dynamics, but
%include vital dynamics in the community model: ${dN_i}/{dt} = \Lambda - \mu_i$, the extended
%Cartesian product of the original SI model with this community model yields the same multi-community
%transmission model as in the previous paragraph.  This illustrates that the model factorization is not
%unique; a given process can be represented as a product of models in different ways.

Other elaborations of the epidemic model can be combined with regional models in the same way, 
including the SIS process (used, for instance, to model gonorrhea (e.g. \cite{lajmanovich-yorke76}), and
more recently to model infectious trachoma \cite{lietman-gebre-ayele2011}), the SIR model, 
more complex variants (e.g., 
\cite{cvjetanovic-grab-uemura1978,hethcote-stech-vandendriessche1981,lechat-mission-vellut1974}
out of a vast literature), or even models featuring vector-borne transmission (e.g. \cite{ross1916,ross1917,bailey1982,smith-battle-hay2012}).
Useful factor models can include regional models with transportation,
host genetics \cite{anderson-may1982,anderson-may1983,anderson-may1991}, gender, vaccination status, 
multiple risk groups (e.g., high and low risk of infection), or the presence of a second infectious agent.

\subsection{Compartmental aging}
Age-structured models are frequently used in analysis of disease transmission to reflect changes in susceptibility,
frequency of complications, or mixing patterns which depend on age.  Compartmental model product structure can
easily reflect these features, as we illustrate in the following example.  Consider the standard SIR model 
to be the first factor model:
\begin{dgroup*}
\begin{dmath*}
\frac{dS}{dt} =  - \beta S I
\end{dmath*} \begin{dmath*}
\frac{dI}{dt} = \beta S I - \gamma I 
\end{dmath*} \begin{dmath*}
\frac{dR}{dt} = \gamma I,
\end{dmath*}
\end{dgroup*}
where $\beta$ and $\gamma$ are transmission and recovery rates
as above.

We then use the following compartmental aging process as the second factor model:
\begin{dgroup*}
\begin{dmath*}
\frac{dA_0}{dt} = \Lambda - \alpha A_0 - \mu_0 A_0
\end{dmath*} \begin{dmath*}
\frac{dA_1}{dt} = \alpha A_0 - \alpha A_1 - \mu_1 A_1
\end{dmath*} \begin{dmath*}
\frac{dA_2}{dt} = \alpha A_1 - \alpha A_2 - \mu_2 A_2
\end{dmath*}
\end{dgroup*}
Here, $\alpha$ is simply the rate of aging (one year per year),
$\Lambda$ is a constant recruitment rate, and constants $\mu_i$
are age-class-specific per capita mortality rates.

The number of age compartments could be chosen to be any positive integer,
in principle.  
Numerically, the use of compartmental aging can yield large stiff systems of equations, but compartmental
aging approximates the use of McKendrick-von Foerster equations for aging.

\begin{figure}
\centering
(a) \\
% WMD file aging-SIR.boxes.crop.pdf
\includegraphics{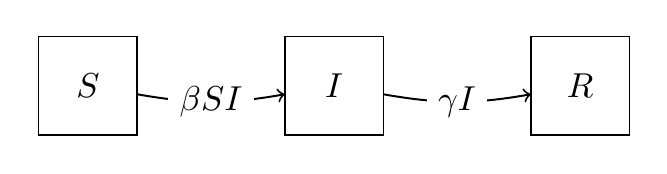}
 \\
(b) \\
% WMD file aging-A.boxes.crop.pdf
\includegraphics{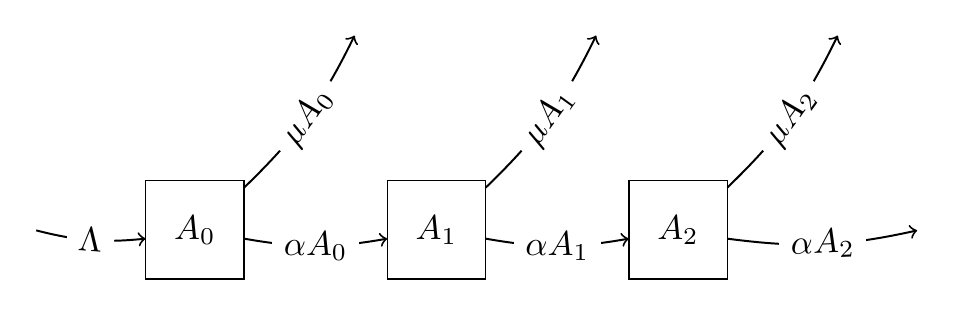}
 \\
(c) \\
% WMD file aging.boxes.crop.pdf
\includegraphics{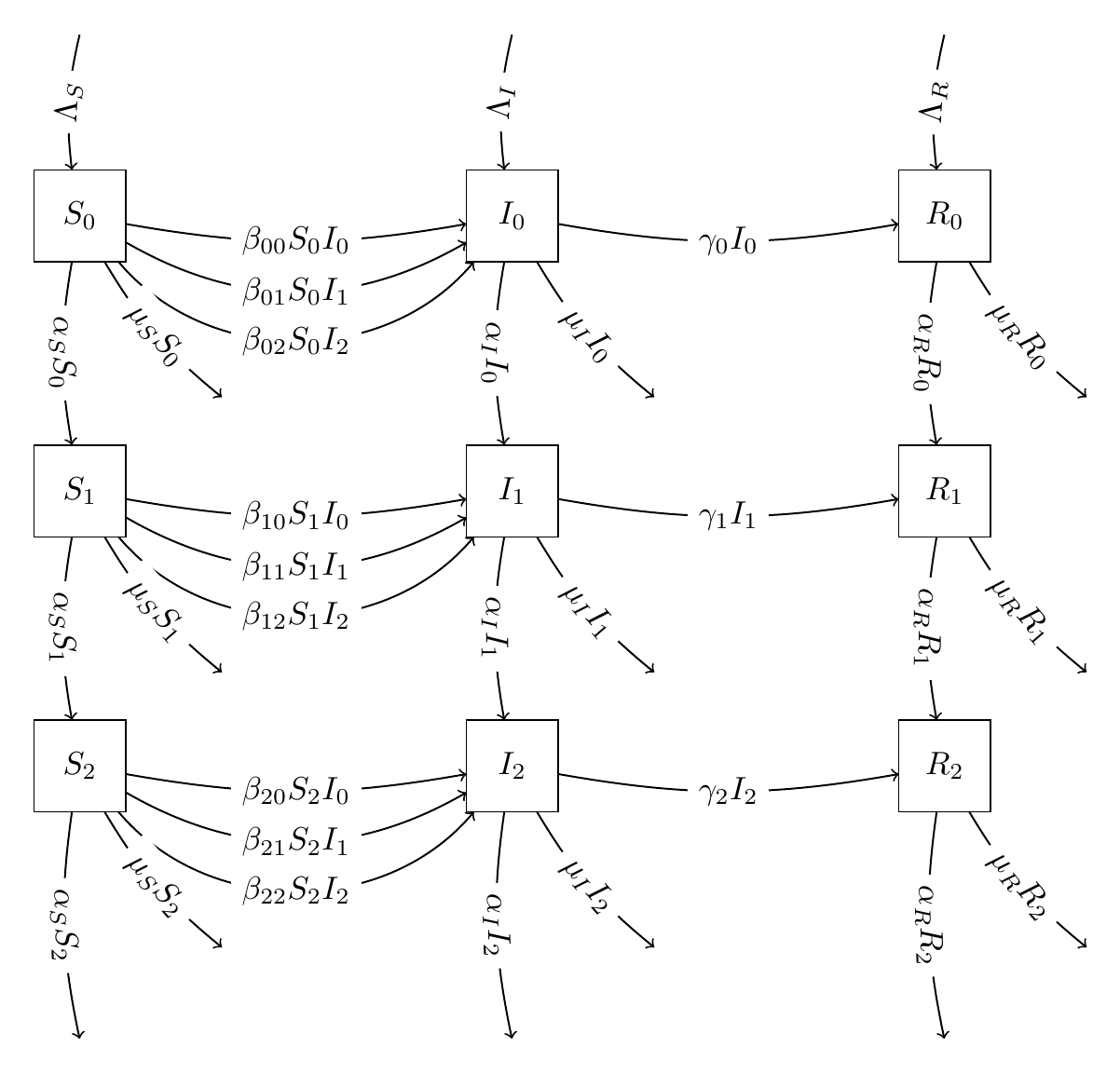}

\caption{ \label{fig:aging}
(a) SIR model, (b) age structure model, and
(c) product of SIR model with age structure model.
%Product (c) of SIR model (a) with age-structure model (b).
}
\end{figure}

These two models and their Cartesian product model are shown in
\hl{Figure \mbox{\ref{fig:aging}}(a),~(b),~and~(c)}.
In the product model, inflow term $\Lambda_I$ represents vertical
transmission while $\Lambda_R$ would represent individuals
immune at birth. Either of these rates can be set to zero
for specific applications.

We note that in this example, we have built recruitment and mortality
into the aging model, while in the previous section we included them
in the transmission model.
There is flexibility in where to include these demographic processes,
depending on what subscripts one wishes to have attached to their rates
in the product model. If needed they can even be included in multiple
factor models and assigned to constant values including zero as appropriate
in the product.

\subsection{Risk-stratified STI model}
Sexual behavior is highly heterogeneous, with some individuals having far more partners per unit time than others.  Moreover,
such individuals may preferentially mix with similar individuals.  The epidemiological role of a relatively small group
of highly active people in
transmission of a sexually transmitted infection (STI) was explored
in a mathematical model of gonorrhea \cite{hethcote-yorke1984},
and similar approaches
were used in HIV modeling \cite{anderson-gupta-ng1989}. 

Consider the following simple example.  Suppose that we begin with the factor model
\begin{dgroup*}
\begin{dmath*}
\frac{dS}{dt} = \Lambda - \beta c p S \frac{I}{N} - \mu S
\end{dmath*} \begin{dmath*}
\frac{dI}{dt} = \beta c p S \frac{I}{N} - \gamma I - \mu I
\end{dmath*}
\end{dgroup*}
representing disease transmission in a population of MSM (men who have sex with men) \cite{anderson-may1991}.
Here $\beta$ represents the transmission probability per partnership,
$c$ equals a susceptible individual's rate of acquiring new sexual partners,
and $p$ is the probability that a susceptible individual's partner
is chosen from a specific population of infectives
(in the basic factor model, there is only one population $I$ of infectives,
and there $p$ is one,
but these probabilities will be nontrivial in the product model,
in which there are multiple infective populations).
Here $\gamma$ is disease-specific per capita mortality,
$\Lambda$ is a constant inflow rate of susceptibles,
and $\mu$ is the disease-independent mortality rate.
We will multiply this model by a second factor model in which the
population is divided into a high risk group A and a low risk group B,
with transition rates $\rho$ and $\sigma$ between them:
\begin{dgroup*}
\begin{dmath*}
\frac{dA}{dt} = - \rho A + \sigma B
\end{dmath*} \begin{dmath*}
\frac{dB}{dt} = \rho A - \sigma B
\end{dmath*}
\end{dgroup*}

\begin{figure}
\centering
% WMD file risk.boxes.crop.pdf
\includegraphics{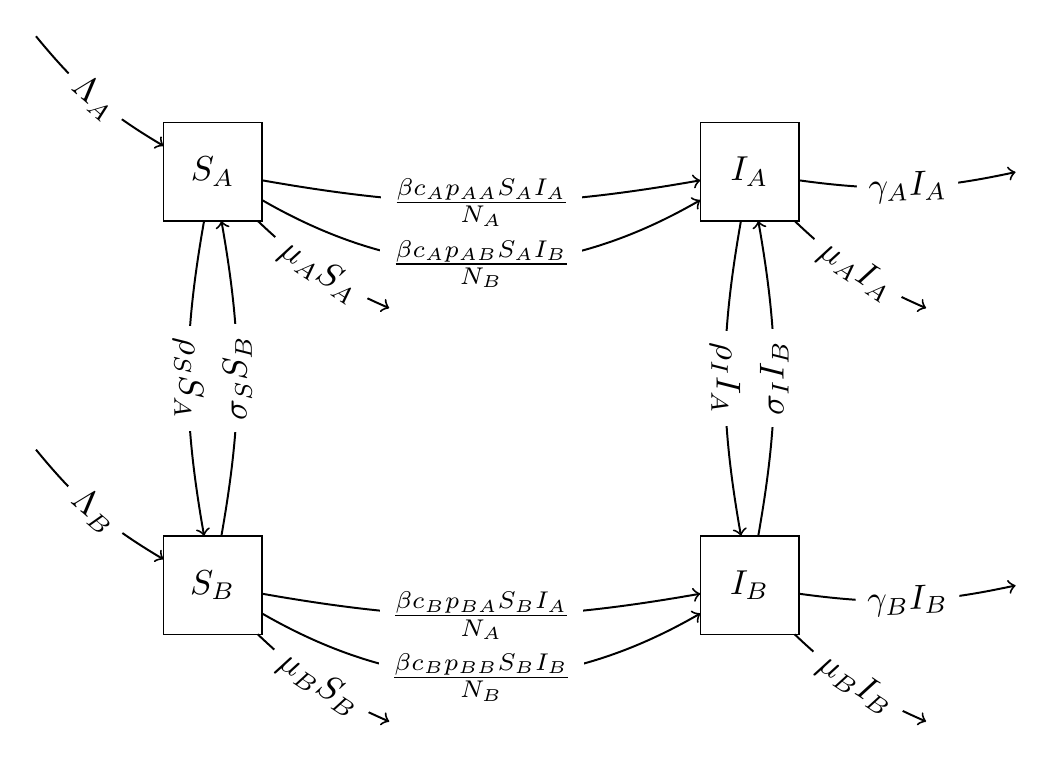}

\caption{ \label{fig:risk}
Product of SI model with risk-structure model.
}
\end{figure}

The product model is shown in \hl{Figure~\mbox{\ref{fig:risk}}}.
This model can then represent the presence of a high-risk
core group with a higher rate $c_A$ of acquiring partners than the other,
as well as nonrandom mixing between the groups, expressed by the
probabilities $p_{AA}$, $p_{AB}$, etc.
Because the mixing probabilities $p_{AA}$ and $p_{AB}$ for a
susceptible individual in group $A$ must sum to one,
and likewise $p_{BA}$ and $p_{BB}$,
we could replace $p_{AB}$ and $p_{BB}$ by $1-p_{AA}$ and $1-p_{BA}$,
but it is not necessary to do so.
As formulated, this system keeps the biologically distinct roles
of $c$ and $p$ separate, although in some circumstances it may
be desirable to combine them,
while respecting the constraint that
$(S_A+I_A) c_A p_{AB}=(S_B+I_B) c_B p_{BA}$
\cite{jacquez1988modeling}.
However, one may desire to have the quantities $p$ be functions
of the state variables, reflecting that partner choice probabilities
may depend on the dynamically varying group sizes
\cite{hethcote-proportionate96,anderson-gupta-ng1989},
in which case it is advantageous to retain them as separate parameters
so that they can be replaced by more complex expressions straightforwardly.

\subsection{Gender in STI models}
\begin{figure}
\centering
% WMD file gender.boxes.crop.pdf
\includegraphics{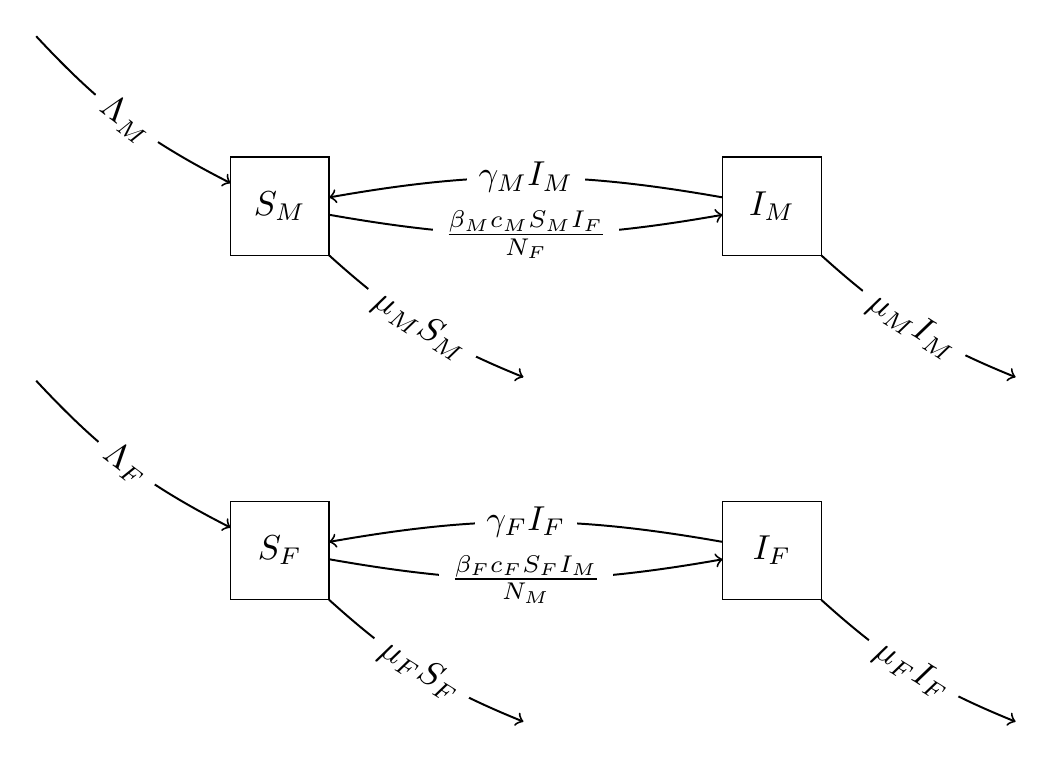}

\caption{ \label{fig:gender}
Product of SI model with static two-gender model.
}
\end{figure}

Modeling heterosexual transmission of an STI may proceed by dividing the population into males and females.  Such a model
can be developed along lines very similar to the risk model in the previous section.  We may begin with a similar
transmission factor model, here shown as an SIS process:
\begin{dgroup*}
\begin{dmath*}
\frac{dS}{dt} = \Lambda - \beta c p S \frac{I}{N} - \mu S + \gamma I
\end{dmath*}
\begin{dmath*}
\frac{dI}{dt} = \beta c p S \frac{I}{N} - \gamma I - \mu I .
\end{dmath*}
\end{dgroup*}
The second factor model will be simply
\[
\frac{dF}{dt} = \frac{dM}{dt} = 0 ,
\]
where we assume no transitions from male to female or vice versa.  
In constructing the product model, we incorporate the assumption of
heterosexual-only transmission by defining the partner-choice probabilities
$p_{ij}$ to be one for opposite-gender combinations ($p_{FM}$, $p_{MF}$)
and zero for the same-gender combinations.
The product model is then
\begin{dgroup*}
\begin{dmath*}
\frac{dS_F}{dt} = \Lambda_F - \beta_{F} c_F \frac{I_M}{N_M} S_F - \mu_F S_F + \gamma_F I_F
\end{dmath*}
\begin{dmath*}
\frac{dS_M}{dt} = \Lambda_M - \beta_{M} c_M \frac{I_F}{N_F} S_M - \mu_M S_M + \gamma_M I_M
\end{dmath*}
\begin{dmath*}
\frac{dI_F}{dt} = \beta_F c_F \frac{I_M}{N_M} S_F - \mu_F I_F - \gamma_F I_F
\end{dmath*}
\begin{dmath*}
\frac{dI_M}{dt} = \beta_M c_M \frac{I_F}{N_F} S_M - \mu_M I_M - \gamma_M I_M,
\end{dmath*}
\end{dgroup*}
as seen in \hl{Figure~\mbox{\ref{fig:gender}}}.

\subsection{Interacting transmission of leprosy and tuberculosis}

\begin{figure}
\centering
% WMD file tuberculosis.boxes.crop.pdf
\includegraphics{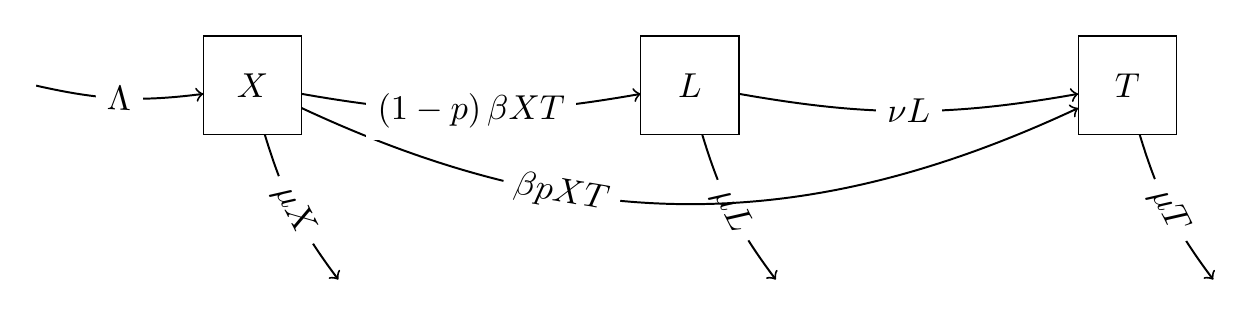}

\caption{ \label{fig:TB}
Compartmental model of tuberculosis transmission
}
\end{figure}

The extended Cartesian product of compartmental models can be applied to problems involving
two separate infectious disease processes.
In the joint leprosy-tuberculosis model appearing in \cite{lietman-porco-blower97}, the
epidemiological effects of cross-immunity between two mycobacterial species was
analyzed using a compartmental model.
This model may be represented as a product of two factor models, the first being a simple tuberculosis model based
on susceptible ($X$), latent TB ($L$), and active tuberculosis ($T$):
\begin{dgroup*}
\begin{dmath*}
\frac{dX}{dt} = \Lambda - \mu X - \beta X T
\end{dmath*}\begin{dmath*}
\frac{dL}{dt} = (1-p)\beta X T - \mu L - \nu L
\end{dmath*}\begin{dmath*}
\frac{dT}{dt} = p\beta X T + \nu L - \mu T  ,
\end{dmath*}
\end{dgroup*}
where $\Lambda$ is a recruitment rate, $\mu$ is an overall mortality rate, $\nu$ is a rate of progression
of latent tuberculosis to active disease, $\beta$ is a transmission coefficient (hazard rate per infective) ($\beta_T$ in 
the paper), and $p$
the probability a newly infected individual will develop active tuberculosis rapidly instead of becoming latently infected with
tuberculosis (\hl{Figure~\mbox{\ref{fig:TB}}}).

\begin{figure}
\centering
% WMD file leprosy.boxes.crop.pdf
\includegraphics{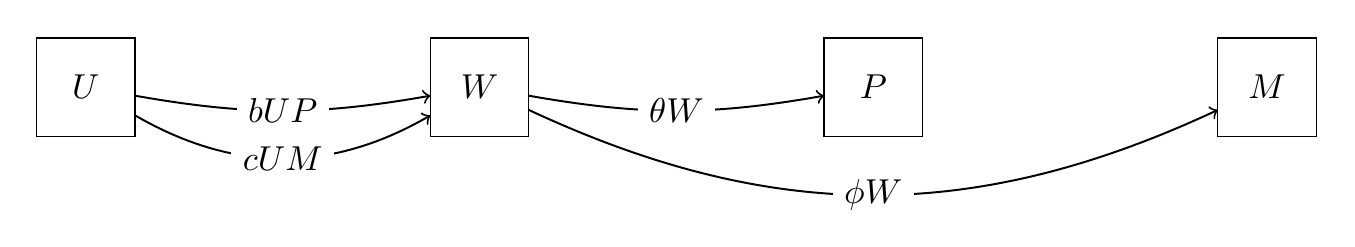}

\caption{ \label{fig:leprosy}
Compartmental model of leprosy transmission
}
\end{figure}

The second factor model represents the progression of leprosy
from susceptible $U$, to latent infection with leprosy ($W$),
and to multibacillary disease ($M$) or paucibacillary disease ($P$).
The leprosy factor model is then (\hl{Figure~\mbox{\ref{fig:leprosy}}})
\begin{dgroup*}
\begin{dmath*}
\frac{dU}{dt} = -(b P + c M) U
\end{dmath*}\begin{dmath*}
\frac{dW}{dt} = (b P + c M) U - (\theta + \phi) W
\end{dmath*}\begin{dmath*}
\frac{dP}{dt} = \theta W
\end{dmath*}\begin{dmath*}
\frac{dM}{dt} = \phi W 
\end{dmath*}
\end{dgroup*}
where here $b$ is the transmission coefficient for paucibacillary leprosy ($\beta_P$ in the paper), $c$ is the transmission coefficient
for lepromatous leprosy ($\beta_M$ in the paper), $\theta$ is the rate at which latently infected individuals develop paucibacillary disease ($\nu_P$ in
the paper), and
$\phi$ is the rate at which latently infected individuals develop multibacillary disease ($\nu_M$ in the paper).

\begin{figure}
\centering
% WMD file leprotb.boxes.crop.pdf
\includegraphics{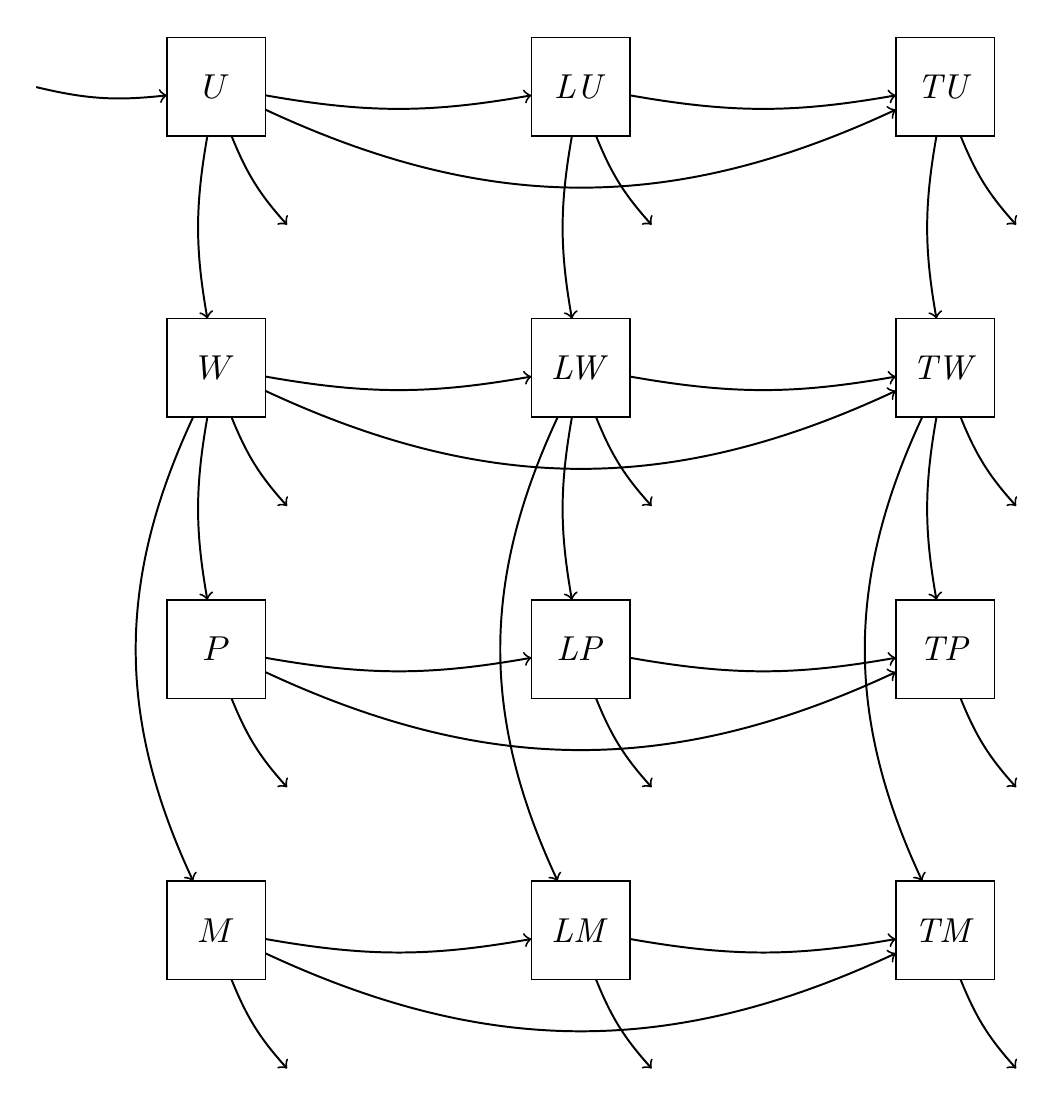}

\caption{ \label{fig:leprosy-TB}
Cartesian product of leprosy (Figure~\ref{fig:leprosy}) and
tuberculosis (Figure~\ref{fig:TB}) models, describing
interaction of the two transmission processes.
Multiple arrows and labels are suppressed for legibility.
}
\end{figure}

The product model (\hl{Figure~\mbox{\ref{fig:leprosy-TB}}})
represents the epidemiological interference of the two closely
related mycobacterial infections.
Individuals latently infected with one may have partial immunity
against the other.
This product structure could be applied to other settings
such as HIV-TB interactions (e.g.\ \cite{porco2001amplification}).

\section{Strong products}

The extended Cartesian product is too
restrictive when constructing models of multiple diseases.  For instance,
one may be infected by two pathogens during a single
encounter with a dually infected person.  Thus, it may be necessary to
allow individuals to 
proceed to dual infection directly from the susceptible class without
passing through the singly infected states.  The extended
Cartesian product defined earlier does not permit this possibility.

Just as the Cartesian product of graphs can be
extended to a strong product of graphs, an analogous strong product
is possible for products of compartmental models.
As we will show below, a strong product of compartmental
models will permit derivation of multi-strain or multi-disease
models featuring simultaneous transmission.

As an example, consider the following simple SIS epidemic model,
which we might apply to transmission of 
{\it Chlamydia trachomatis}, the etiologic agent of trachoma (a
blinding disease) \cite{gao-lietman-dong2016,lietman-porco-dawson99}.
In principle, multiple strains of the trachoma
agent can circulate \cite{kari-whitmire-carlson2008}.
Consider a model of a single strain, in which $S$ is the number of 
susceptibles, and $I$ the number of infectives:
\begin{dgroup*}
\begin{dmath*}
\frac{dS}{dt} = -\beta S \frac{I}{N} + \gamma I
\end{dmath*}
\begin{dmath*}
\frac{dI}{dt} = \beta S \frac{I}{N} - \gamma I
\end{dmath*}
\end{dgroup*}
where $\beta$ is a transmission coefficient, $\gamma$ is a recovery rate,
and $N=S+I$ is the total population.

We can construct a multistrain model with partial cross-immunity
\cite{dietz79,keeling} using a suitably defined strong product,
defined as follows.  Let $A$ and $B$ be models, with states labeled
$A_i$ and $B_j$ respectively.
The vertices of the strong product model $A\,\boxtimes\,B$ are,
as in the previously defined products,
the Cartesian product of the vertex sets of the factor models.
Every arc of each factor model gives rise to one or more arcs
within each level of the product model, as in the other products,
one for each interaction with product compartments.
There are also additional, diagonal arcs in the product model
that cross levels, representing more than one of the factor
model's transitions taking place simultaneously.

\begin{figure}
\centering
% WMD file SIH-strongg.boxes.crop.pdf
\includegraphics{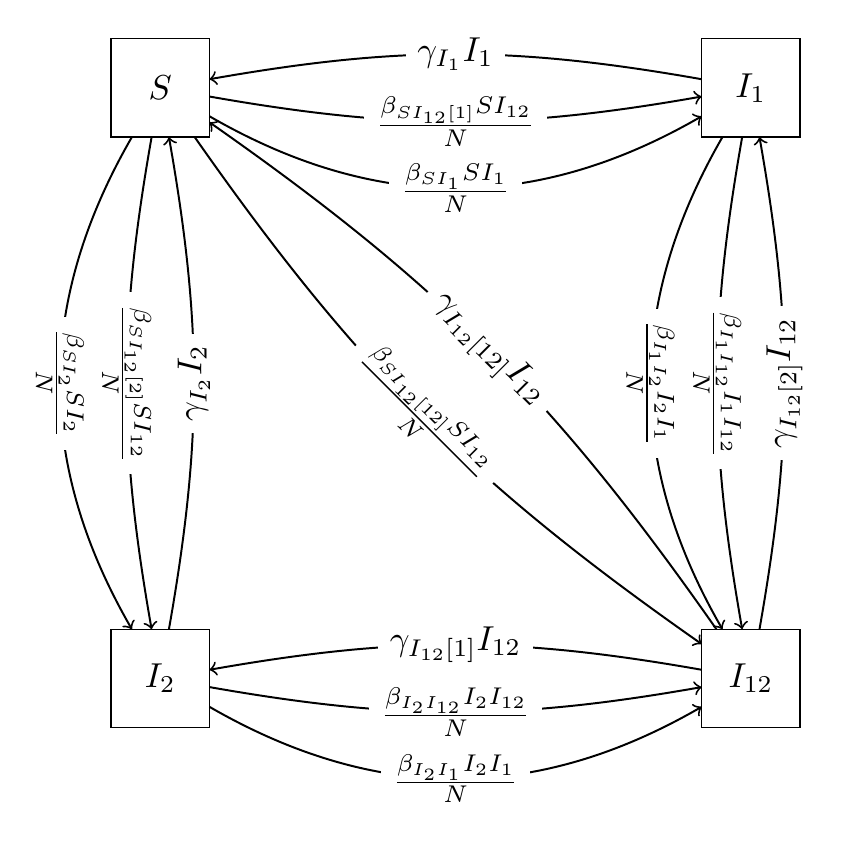}

\caption{ \label{fig:strong}
Strong product of two strains' SIS dynamics.
}
\end{figure}

In \hl{Figure~\mbox{\ref{fig:strong}}} we present the strong
product of the above SIS model with itself.

The arcs of the Cartesian product of models are present,
representing infection of an individual by either strain
1 or 2,
but there are additional arcs as well, including
the diagonal transition from $S$ to $I_{12}$
representing simultaneous transmission of both strains
1 and 2 to a fully susceptible individual in a single
encounter with an individual carrying both strains.

Unlike the Cartesian product, here a single interaction between
compartments can manifest in multiple transitions.
The interaction between example $S$ and $I_{12}$ can result in
transmission of either or both strains, and these cases
are represented by three arcs in the diagram,
with transmission rates $\beta$ distinguished by brackets.

A formal definition of the strong product of compartmental
models, which generates the above example product, is as follows:

\begin{definition}  A strong product of two
compartmental models $A$ and $B$ is a compartmental
model $A\boxtimes B$ whose set of compartments is
the set of ordered pairs $(X_i,Y_j)$ for every compartment $X_i$
of $A$ and $Y_j$ of $B$.
Models $A$ and $B$ are considered to be distinct models for the
purpose of this definition, and each model's transitions are considered distinct
from the other's, even when a product is taken of a model with itself.

For every set $\alpha=\{\alpha_1,\ldots,\alpha_p\}$ of factor models' transitions,
belonging to distinct factor models,
each with
source compartment $X^{\alpha_i}_s$, target compartment $X^{\alpha_i}_t$,
and per capita transition rate
$f^{\alpha_i}(X^{\alpha_i}_s,Z^{\alpha_i}_1,\ldots,Z^{\alpha_i}_{k_i})$,
the product model includes all arcs of the form
\[ x_s \xrightarrow{f^{\alpha_1,\ldots,\alpha_p}_{i,\ldots}(x_s,z_{1},\ldots,z_{k})} x_t \]
where $x_s$ ranges over the compartments of the product model
that descend from all the factor vertices
$\{X^{\alpha_1}_s,\ldots,X^{\alpha_p}_s\}$,
each $z_j$ ranges over the product compartments
that descend from all vertices in the set $\{Z^{\alpha_i}_j\}$,
and $x_t$ descends from all of $\{X^{\alpha_i}_t\}$ and
is otherwise on all the same levels as $x_s$.
The arcs of the product model are only those generated by the
above definition.
As previously, the subscripts $i,\ldots$ to the rate function $f$
distinguish the different product arcs by indicating the levels to which
all the function's arguments belong.
\end{definition}

In our SI example, we have defined the rate functions $f$ to
produce appropriate products of
transmission ($\beta$) and recovery ($\gamma$)
transitions, with distinct but compact subscripts,
and to omit transitions in which transmission of one strain
occurs simultaneously with recovery from the other one.

\section{Exploration of a family of models}

%In this section we show how using this formalism and its software
%implementation makes it tractable to generate a zoo of models
%representing addition of variable amounts of detail to a
%model transmission process.

% @@ get rid of zeta
% @@ don't do per-capita rates in diagrams

\begin{figure}
\[
\begin{array}{ll}
\textbf{A.}
% WMD file c-transmission.boxes.crop.pdf
\includegraphics{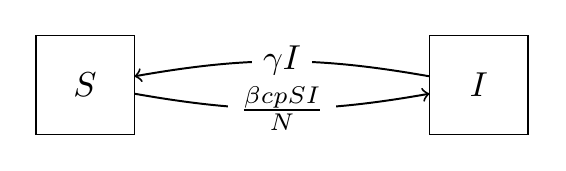}

&
\textbf{B.}
% WMD file c-gender.boxes.crop.pdf
\includegraphics{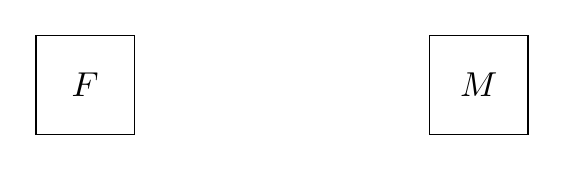}

\\
\textbf{C.}
% WMD file c-risk.boxes.0.8.crop.pdf
\includegraphics{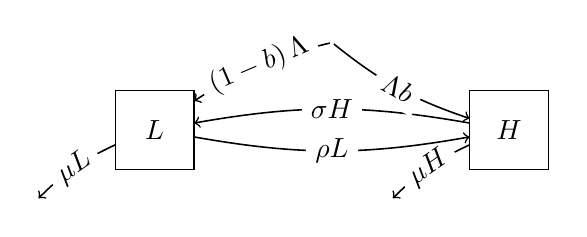}

&
\textbf{D.}
% WMD file c-exp.boxes.crop.pdf
\includegraphics{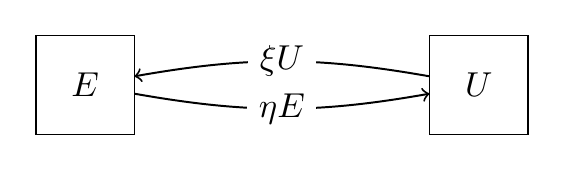}

\end{array}
\]
\caption{\label{fig:gonorrhea-factors}
Component models for gonorrhea process:
\textbf{A.} Transmission model, a classic SIS process;
\textbf{B.} Gender model, a male-female binary system, with
the assumptions that nonbinary proportions and transition rates are low;
\textbf{C.} Risk model, consisting of high and low risk groups;
\textbf{D.} Exposure model, consisting of groups unexposed and exposed
to screening.}
\end{figure}

In this section we illustrate the use of the extended Cartesian product
in model development and exploration,
using a model of targeted screening for gonorrhea as a
simple example.
Such a model can be expressed using four components:
a natural history model,
a partitioning of the population by gender,
a division into low and high risk groups,
and a process of screening of individuals
(Figure~\ref{fig:gonorrhea-factors}).
For the natural history model we will use the simple SIS process
as in \cite{lajmanovich-yorke76} for illustration,
while recognizing that for some STIs a more complex transmission
model may be needed, for example to reflect partial immunity
\cite{brunham2005unexpected}.
Because within- and between-gender transmission can vary greatly,
we include a division of the model population by gender,
with the assumption that rates of gender transition and
proportions of non-binary individuals are small in comparison
to the model dynamics.
We include a high and low risk group as in \cite{hethcote-yorke1984},
with transitions between the risk groups,
and finally we include an exposure model tracking the individuals
exposed to a control measure such as frequent screening \cite{lee2016sexually}.

\begin{figure}
\centering
% WMD file c-2222.boxes.0.7.crop.pdf
\includegraphics{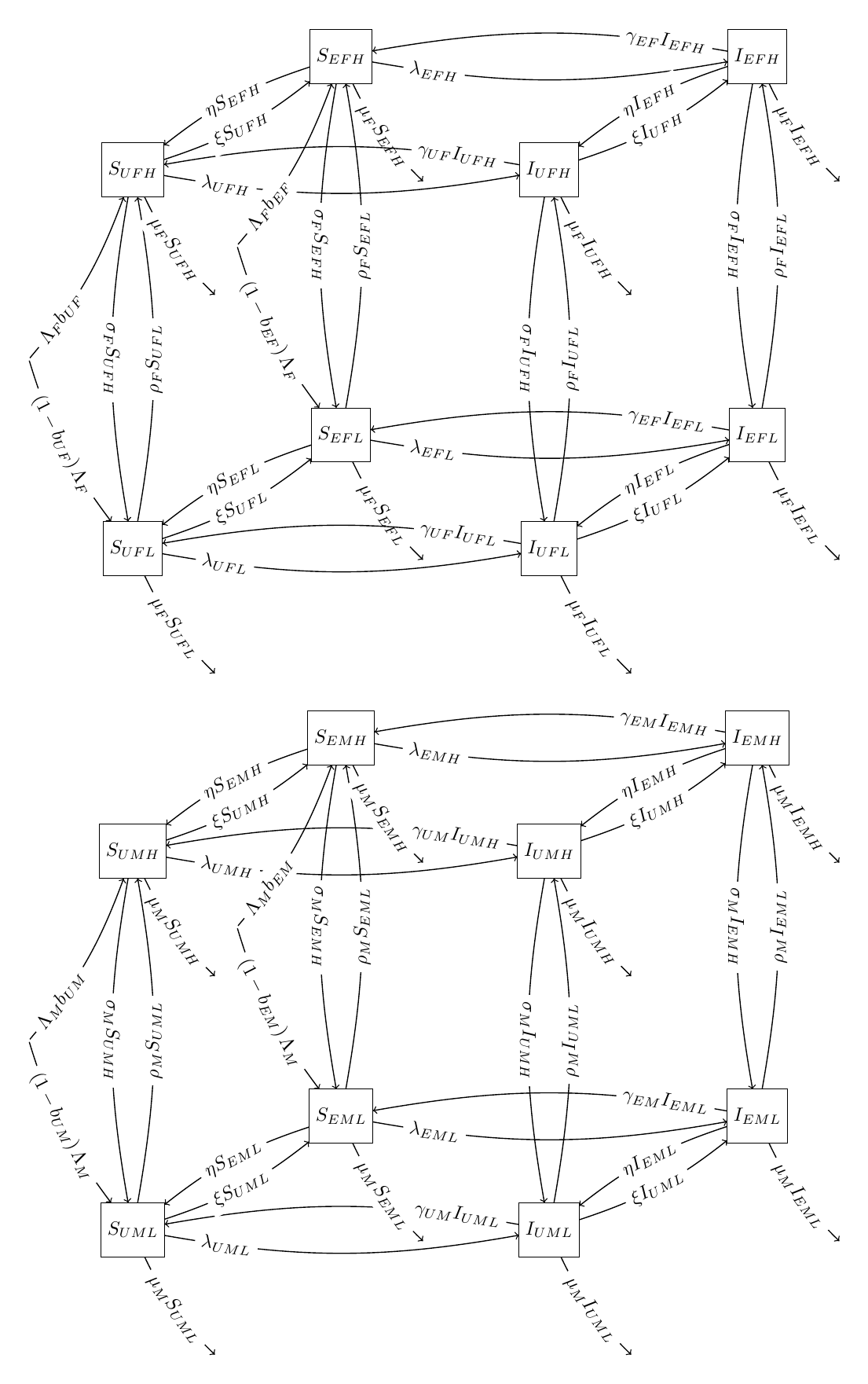}

\caption{\label{fig:gonorrhea-product}
Four-level product model of gonorrhea transmission
with stratification into gender, risk, and exposure categories.
Transmission rate is abbreviated here for readability:
$\lambda_{abc}=S_{abc}c_{bc}\sum_{def}\beta_{be}p_{bcef}I_{def}$,
where $a,b,c$ and likewise $d,e,f$ range over exposure, gender,
and risk groups respectively.}
\end{figure}

The product of these four models
(constructed by extending the above definition of the extended Cartesian
product of two models, or by taking a product of products)
has sixteen compartments and describes a process of transmission
with rates affected by the genders, risk group membership,
and exposure status of both susceptibles and infectives
(Figure~\ref{fig:gonorrhea-product}).
The product structure naturally generates a process that includes
both homosexual and heterosexual transmission.
% @@ fix lambdas in figure caption
As drawn here, the effect of the screening program is expressed
by changes in the removal rate $\gamma$
such that screened individuals are removed from the infective state
more quickly than those who are not screened.
% @@ put it on gamma
% @@ take it off p
% @@ put on github

\begin{figure}
\[
\begin{array}{ll}
\textbf{A.}
% WMD file c-2211.boxes.crop.pdf
\includegraphics{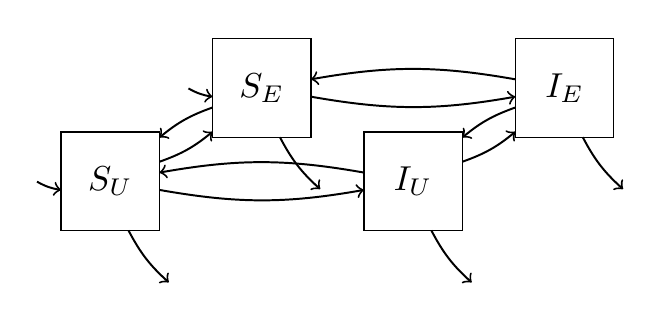}

&
\textbf{B.}
% WMD file c-2121.boxes.crop.pdf
\includegraphics{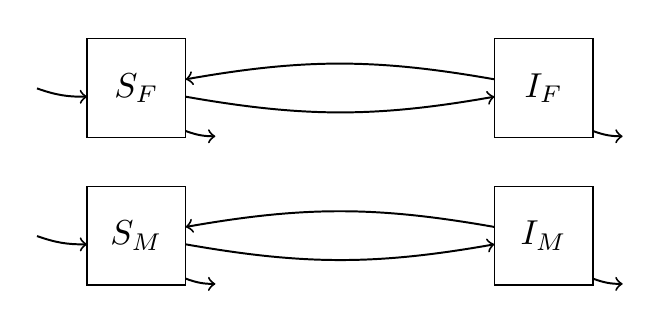}

\\
\textbf{C.}
% WMD file c-2112.boxes.crop.pdf
\includegraphics{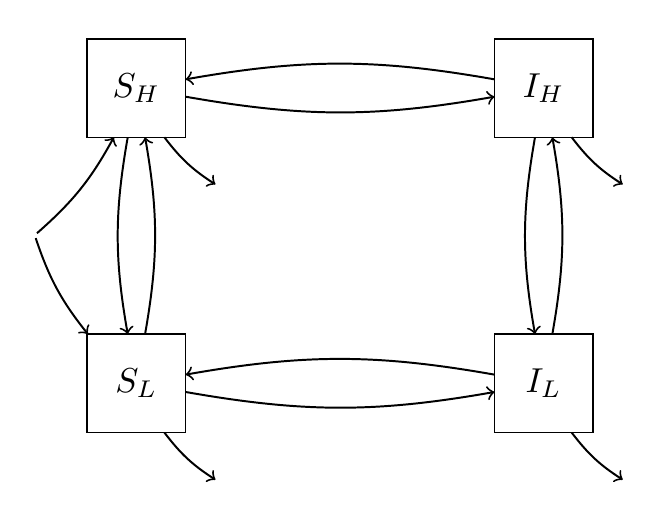}

&
\textbf{D.}
% WMD file c-2221.boxes.crop.pdf
\includegraphics{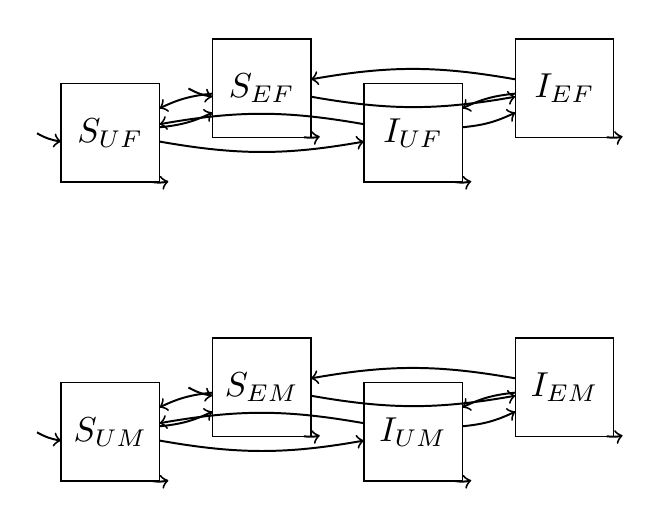}

\\
\textbf{E.}
% WMD file c-2212.boxes.crop.pdf
\includegraphics{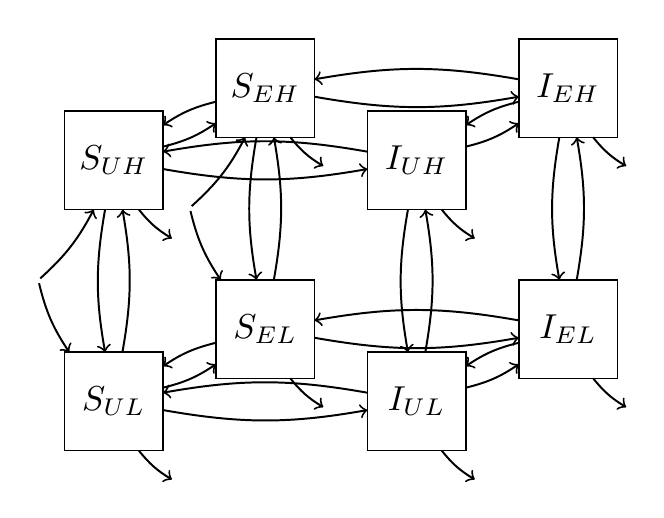}

&
\textbf{F.}
% WMD file c-2122.boxes.crop.pdf
\includegraphics{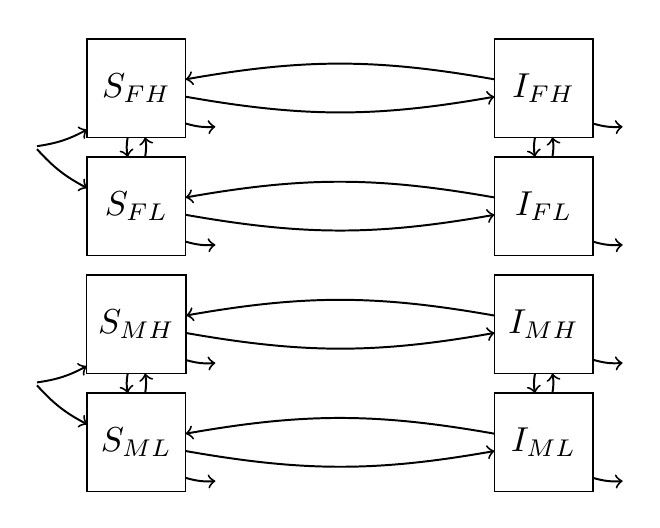}

\end{array}
\]
\caption{\label{fig:gonorrhea-marginals}
Some candidate models for explanation of recorded transmission
dynamics:
\textbf{A.} Transmission with exposure only;
\textbf{B.} with gender only;
\textbf{C.} with risk groups only;
\textbf{D.} with exposure and gender;
\textbf{E.} with exposure and risk groups;
\textbf{F.} with gender and risk groups.
}
\end{figure}

Using the extended Cartesian product definition of this model,
it is straightforward to generate partial products using subsets
of the set of four factor models shown in Figure~\ref{fig:gonorrhea-factors},
yielding a spectrum of models of intermediate complexity
(Figure~\ref{fig:gonorrhea-marginals}),
which can be evaluated on their ability to fit observed data.
Methods to evaluate the goodness of fit of such a model
to data might include least squares (e.g.\ \cite{hethcote2013modeling})
or likelihood methods (e.g.\ \cite{blok2017forecasting}).

\begin{figure}
\centering
% WMD file c-2223.boxes.0.8.crop.pdf
\includegraphics{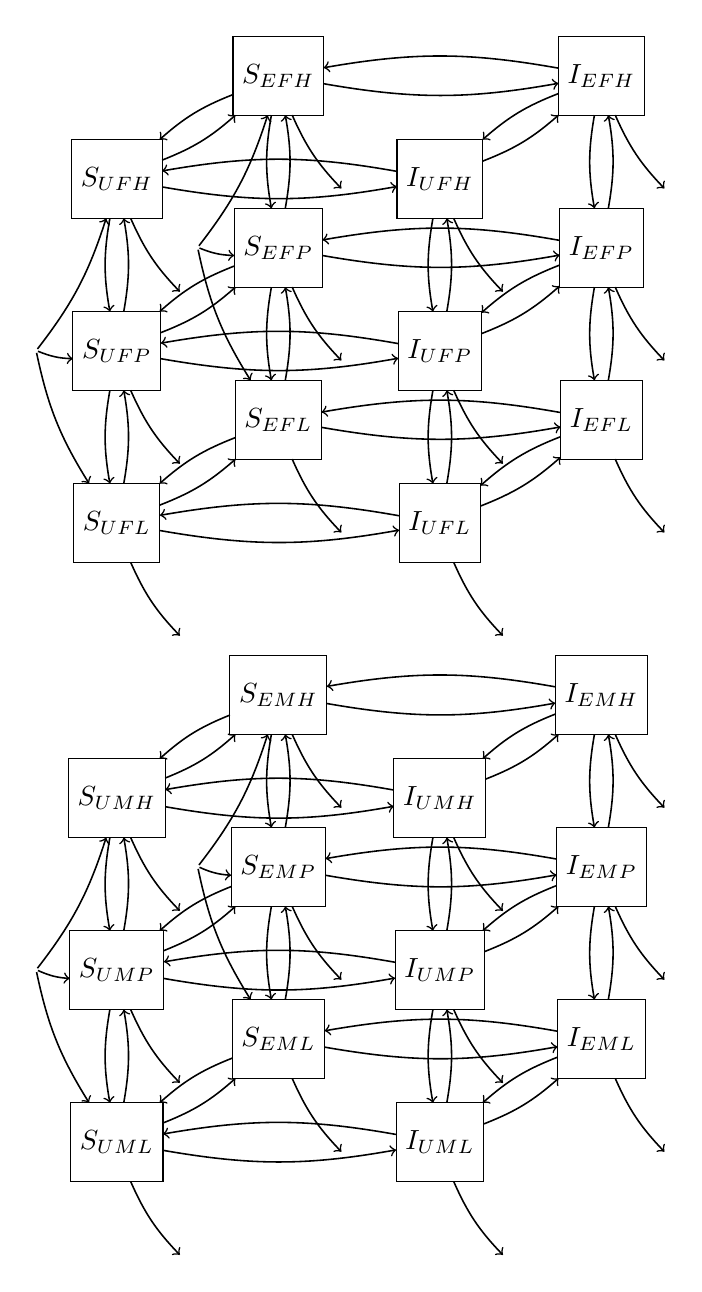}

\caption{\label{fig:gonorrhea-2223}
Transmission dynamics with three risk groups (Low, Partial, and High),
together with gender and exposure categories.}
\end{figure}

More importantly, it is also straightforward using this formulation
to generate models with greater detail, for example by using more
than two risk groups (Figure~\ref{fig:gonorrhea-2223}).
In this way, models with arbitrarily large numbers of risk groups
can be straightforwardly and systematically evaluated for
goodness of fit to find the best description
of the true process available in this framework,
a process which can not be undertaken without an automated model
generation framework of this sort.

\section{Discussion}
Products of compartmental models, defined as straightforward
generalizations of graph products, represent useful operations in
developing epidemiological models.  Similar mathematical
structures arise from addition of 
age structure, gender, geographic differences, or other
forms of heterogeneity to an epidemic model.  Such
similarities reveal the presence of a ``design pattern''
\cite{alexander1977pattern}
that is captured by the extended Cartesian products we define here.

The products presented in this paper by no means represent
the full range of possible products of models.  For
the products we presented,
the state space of the product model is the Cartesian product
of the state spaces of the factor models.  Some cases may
require only a subset of this---consider an HIV model in which
the infective classes are structured by CD4 count and viral load
classes, which are not relevant to the susceptible classes.  
Such examples can be easily handled by straightforward generalizations
of the products given in this paper.  More complex products
are required when the state space of the product model must
include history (for example, the order in which individuals
were infected by pathogen strains).  

The extended Cartesian product is well suited to
the operation of adding host heterogeneity to
an epidemic model, and so it
may facilitate automated generation of a family of
epidemic models.  Similarly, the extended strong product
is suited to the process of adding pathogen heterogeneity to
an epidemic model.
We have developed software to implement these products.
All models and figures in this paper were
generated by this software, which is freely available
as a module for the Sage mathematics computing system
\cite{worden-boxmodels-github}.
This software enables systematic numerical exploration of
a large family of related models, to automate
evaluation of specific refinements of an epidemic process for relevance
to observed dynamics, and
could form the basis of a systematic approach to
numerical structural sensitivity analysis.

\section{Acknowledgements}

This study was supported by a Models of Infectious Disease Agent Study
(MIDAS) grant from the US NIH/NIGMS to the University of California, San
Francisco (U01GM087728). The Department of Ophthalmology acknowledges
support from Research to Prevent Blindness.

\section{Conflict of Interest Statement}

The authors declare they each have no conflicts of interest.

%====================================================
% Bibliography
%====================================================

\section{References}

%\newpage
\bibliographystyle{elsarticle-num}
\bibliography{box}

%====================================================
% Appendix
%====================================================

\appendix

\section{Graph products}
\label{app:graph-products}

Formally a directed graph is a set of vertices together with
arcs, defined as a set of ordered pairs of vertices.  We consider
graph products for which the vertices of the graph product are
formed from the Cartesian product of the vertices of each
of the factor graphs.  Notationally, if
$S_1$, $S_2$, $\ldots$, $S_n$ are sets of vertices, then their
Cartesian product is the set
$S_1\times S_2\times\ldots\times S_n = \{\,(s_1,s_2,\ldots,s_n)\mid s_1\in S_1, s_2\in S_2,\ldots, s_n\in S_n\,\}$,
with the elements of $S_1\times S_2\times\ldots\times S_n$ being
tuples of elements of the component sets $S_1$, $S_2$, $\ldots$, $S_n$.

For application to mathematical models, the arrows (arcs) will
represent transition rates between states represented by vertices.  Each
arc will therefore require an associated label.  
We allow multiple arcs between the same vertices and thus
we must use multigraphs,
represented
as a set $\{\,(v,w,e)\,\}\subseteq V\times V\times E$. Here,
$V$ is the vertex set of the graph and $E$ is its set of arc
labels.  Each of these tuples is visualized as an arrow from $v$ to
$w$ with label $e$ (which we may denote
$\{\,v\xrightarrow{e} w\,\})$.

The Cartesian product
$G_1\cartprod G_2\cartprod\cdots\cartprod G_n$
of directed graphs
$G_1$, $G_2$, $\ldots$, $G_n$ is a graph whose vertex
set is the Cartesian product $V_1\times V_2\times\ldots\times V_n$
of the vertex sets $V_i$ of each graph $G_i$, and whose arcs
are of the form
$(v_1,v_2,\ldots,v_i,\ldots,v_n)\xrightarrow{e} (v_1,v_2,\ldots,w_i,\ldots,v_n)$,
where the two tuples are identical in all but the $i$'th position,
and where there is an arc connecting $v_i$ to $w_i$ in $G_i$.
In this paper, the products defined will yield the arc 
labels (transition rates) in the product graph.  

For each
vertex in one of the factor models, each vertex in the other model
is replicated.  
If the first factor is a graph with vertices A and B, with an arc from A 
to B, and the second is a graph with vertices 1 and 2 and an arc from 1 to 2, then the
product graph contains vertices which could be denoted A1, A2, B1, and B2.  Each arc
of one graph is replicated for every vertex of the other.  Thus, for example, 
the arc from 1 to 2 in the second factor graph corresponds to an arc from A1 to A2 for
the first vertex of the first model, and also to an arc from B1 to B2 for the second vertex.
Similary, the arc from A to B in the first model corresponds to an arc from A1 to B1 (for arc 1
of the second model), and from A2 to B2 (for the other).

The strong product (or strong Cartesian product) of graphs $G_1$, $\ldots$, $G_n$, written
$G_1\,\boxtimes\,G_2\,\boxtimes\,\cdots\,\boxtimes\,G_n$, is
the graph whose vertex set is the Cartesian product of the graphs'
vertex sets, and which has an arc from
$(v_1,\ldots,v_n)$ to $(w_1,\ldots,w_n)$ if and only if, for
every $i$, either there is an arc from $v_i$ to $w_i$, or
$v_i=w_i$.
The Cartesian product
$G_1\cartprod G_2\cartprod\cdots\cartprod G_n$
of directed graphs
$G_1$, $G_2$, $\ldots$, $G_n$ is a graph whose vertex
set is the Cartesian product $V_1\times V_2\times\ldots\times V_n$
of the vertex sets $V_i$ of each graph $G_i$, and whose arcs
are of the form
$(v_1,v_2,\ldots,v_i,\ldots,v_n)\xrightarrow{e} (v_1,v_2,\ldots,w_i,\ldots,v_n)$,
where the two tuples are identical in all but the $i$'th position,
and where there is an arc connecting $v_i$ to $w_i$ in $G_i$.
%In this paper, the products defined will yield the arc 
%labels (transition rates) in the product model.

\subsection{Adjacency matrices}
\label{app:graph-product-matrices}

The Cartesian product graph can also be defined by
its adjacency matrix.  
Let the adjacency matrices of finite graphs $A$ and $B$, respectively, be
$M_A$ and $M_B$.  Let $I_A$ and $I_B$ be identity matrices of the
same size as $M_A$ and $M_B$ respectively.
The Cartesian product graph $A\cartprod B$ has adjacency matrix
$M_C = M_A \oplus M_B = M_A \otimes I_B + I_A \otimes M_B$,
where $\otimes$ is the Kronecker product.
Writing $M_C'=M_B \oplus M_A$ yields an adjacency matrix for the product graph which is
the same, except for the ordering of the vertices in the product (and the order in
which the Cartesian product of sets of vertices is taken).
More generally, graphs with multiple arcs between two vertices can be defined,
in which case the elements $a_{ji}$ of
the adjacency matrix record the number of directed arcs to $j$ from $i$.
The Cartesian product definition
above can be applied in this case as well.

The strong product of two directed graphs $A$ and $B$
includes more arcs than the Cartesian product\cite{imrich-klavzar2000}.
This product $A\boxtimes B$ has the same vertex set,
the Cartesian product of the factors' vertex sets,
but in addition to the arcs of the Cartesian product graphs,
it also includes all arcs from $(v_s,w_s)$ to $(v_t,w_t)$
where there is an arc from $v_s$ to $v_t$ \emph{and} an arc
from $w_s$ to $w_t$.
Using the same notation as above, the adjacency matrix of
the strong product graph is
$M_D=I_A \otimes M_B + M_A \otimes I_B + M_A \otimes M_B$.

Products of more than two graphs can be represented by
analogous, though more tedious, matrix operations.

\subsection{Linear compartmental models}
\label{app:markov-products}

Suppose we consider a compartmental model with states $X_0$, $X_1$, $\ldots$, $X_k$ 
represented by the first order linear system
\[
\dot{X} = a + M X
\]
where $a$ is a vector of exogenous inflow terms, and $M$ is a transition rate
matrix.  In general, $M$ could contain sink terms;
let $X_1$ represent the
number of individuals in a population with constant recruitment $\Lambda$ and
constant per-capita mortality, so that
example $\dot{X_1} = \Lambda - \mu X_1$.  In this case
$M$ is a $1 \times 1$ matrix, $[-\mu]$.

Consider two continuous time chains, with 
state spaces $X_i^{A}$, $i=1,\ldots,n_A$ and $X_j^{B}$, $j=1,\ldots,n_B$
respectively.  Suppose that the transition rate matrix for each is
given by $M_A$ and $M_B$ respectively (assumed time-independent,
for simplicity).  A
transition rate matrix may contain both positive (inflow) terms, and negative
(outflow) terms.  A transition to $X_k^A$ from state $X_i^A$ at rate $\lambda$
will be represented in $M_A$ by a term in the $i,i$-th element of $M_A$ of $-\lambda$,
and a term in the $k,i$-th element of $M_A$ of $\lambda$.
\begin{dmath*}
M_A = \left[ \begin{array}{ccc}
-\lambda & \mu & \sigma \\
\lambda & -\mu - \theta & 0 \\
0 & \theta & -\sigma \\
\end{array}\right]
\end{dmath*}
%Such a process can be represented by a directed graph, as in
%figure~\ref{fig:Markov-products}, with an arc for each transition
%labeled by its transition rate.
Any transition
rate matrix can be represented as a sum over arcs:
\begin{dmath*}
M_A = \left[ \begin{array}{ccc}
-\lambda & 0 & 0 \\
\lambda & 0 & 0 \\
0 & 0 & 0 \\
\end{array}\right]
 + \left[ \begin{array}{ccc}
0 & \mu  & 0 \\
0 & -\mu & 0 \\
0 & 0 & 0 \\
\end{array}\right]
+ \left[ \begin{array}{ccc}
0 & 0 & 0 \\
0 & - \theta & 0 \\
0 & \theta & 0 \\
\end{array}\right]
+ \left[ \begin{array}{ccc}
0 & 0 & \sigma \\
0 & 0 & 0 \\
0 & 0 & -\sigma \\
\end{array}\right].
\end{dmath*}

\begin{figure}
\centering
(a) % WMD file M_A.crop.pdf
\includegraphics{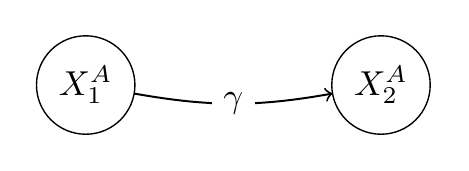}
\\
(b) % WMD file M_B.crop.pdf
\includegraphics{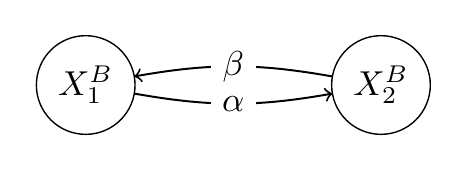}
\\
(c) % WMD file Markov-product.crop.pdf
\includegraphics{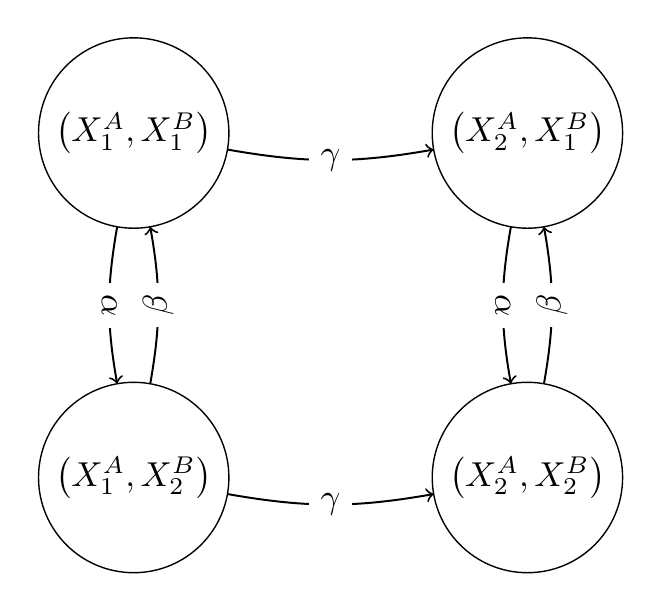}
\\
(d) % WMD file Markov-product-lambda.crop.pdf
\includegraphics{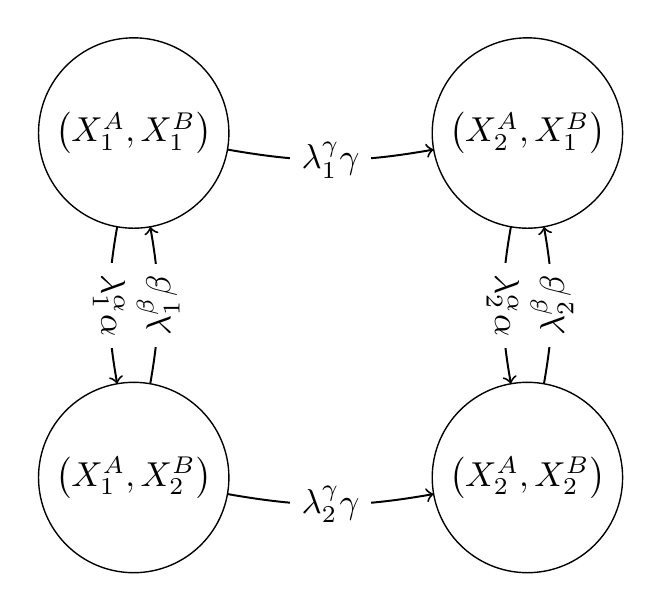}
\caption{ \label{fig:Markov-products}
(a) Diagram of states and transition rates for example Markov model;
(b) diagram for second example Markov model;
(c) diagram of states and transition rates for simple Cartesian product
of models;
(d) diagram of states and transition rates for general Cartesian product
of models, as defined in the text.
}
\end{figure}

A Cartesian product of two Markov chains $A$ and $B$ can then be defined as follows.
We will denote the arcs of model $A$ by $\alpha_A$; to each arc corresponds the matrix $M_A^{\alpha_A}$
with elements 0 everywhere, except for the inflow and outflow represented by that arc, as illustrated
above.
The decomposition of the transition rate matrix $M_A$ by arcs is then $M_A = \sum_{\alpha_A} M_A^{\alpha_A}$.
Similarly, the transition rate matrix $M_B$ of model $B$, decomposed by arcs, is
$M_B = \sum_{\alpha_B} M_B^{\alpha_B}$.  If $I_A$ and $I_B$ are identity matrices of the same dimension as
$M_A$ and $M_B$ respectively, a special case of the Cartesian product can be written
\[
M_C = \sum_{\alpha_A} 
I_B \otimes M_A^{\alpha_A} + \sum_{\alpha_B} M_B^{\alpha_B} \otimes I_A.
\]
\hl{Figure~\mbox{\ref{fig:Markov-products}}} depicts two
such models (\hl{Figure \mbox{\ref{fig:Markov-products}}(a)~and~(b)})
and their product as defined here
(\hl{Figure~\mbox{\ref{fig:Markov-products}(c)}}).
This special case only represents a model in which
the two chains behave completely independently.  

To obtain a more general product, we replace the identity matrices with general diagonal matrices.  
The elements of the diagonal matrix are not assumed identical.
We let $\Lambda_A^{\alpha_B}$ be a such a diagonal
matrix of the same dimension as $M_A$, representing
a scaling matrix for each arc of model $B$ for every state of model $A$.  Similarly, $\Lambda_B^{\alpha_A}$,
a matrix of the same dimension as $M_B$, 
is a scaling matrix for each arc of model $A$ for every state of model $B$.  A more general product is then
expressed by
\[
M_C = \sum_{\alpha_A} \Lambda_B^{\alpha_A} \otimes M_A^{\alpha_A} + \sum_{\alpha_B} M_B^{\alpha_B} \otimes \Lambda_A^{\alpha_B}.
\]
producing a product process like the one pictured in
\hl{Figure~\mbox{\ref{fig:Markov-products}}(d)}.

Note that a given arc connecting two vertices may, in applications, represent two separate processes.
For instance, we may have a compartment representing live individuals and another dead, and wish to 
model the rate of death due to two causes, say $\mu_1$ and $\mu_2$.  Canonically, we may represent the total transition
from live to dead as a single arc with rate $\mu_1+\mu_2$ (assuming independent competing risks), but
in the decomposition above, if we represent the transition by two separate arcs, the Cartesian
product formula may be applied in the same way.

Let us use the notation $z\propto Z$ to indicate that
compartment $z$ descends from factor compartment $Z$.
The transition matrix for the extended Cartesian product model is
\begin{dmath*}
M_C = \sum_{\alpha} \left[ \sum_{z_1\propto Z^{\alpha}_1} \cdots \sum_{z_k\propto Z^{\alpha}_k} \Lambda^{\alpha,z_1,\ldots,z_k}_B \otimes M_A^{\alpha}(z_1,\ldots,z_k) \right] +
\sum_{\alpha} \left[ \sum_{z_1\propto Z^{\alpha}_1} \cdots \sum_{z_k\propto Z^{\alpha}_k} M_B^{\alpha}(z_1,\ldots,z_k) \otimes \Lambda^{\alpha,z_1,\ldots,z_k}_A \right]
\end{dmath*}
where $k$ is considered to depend on $\alpha$.

\end{document}